\documentclass[10pt,journal,compsoc,onecolumn,final]{IEEEtran}

\usepackage[cmex10]{amsmath}
\usepackage{amssymb}
\usepackage{amsthm}
\usepackage{booktabs}
\ifCLASSOPTIONcompsoc
  \usepackage[nocompress]{cite}
\else
  \usepackage{cite}
\fi
\usepackage{color}
\definecolor{darkblue}{cmyk}{0.83,0.89,0.0,0.43}
\usepackage[T1]{fontenc}
\ifCLASSINFOpdf
  \usepackage[pdftex]{graphicx}
\else
  \usepackage[dvips]{graphicx}
\fi

\usepackage[colorlinks,linkcolor=darkblue,urlcolor=blue,citecolor=darkblue,hyperindex,plainpages=false]{hyperref} 

\usepackage{algorithm} 
\usepackage{algorithmicx}
\usepackage{algpseudocode}

\usepackage[latin1]{inputenc}
\ifCLASSOPTIONcompsoc
  \usepackage[caption=false,font=footnotesize,labelfont=sf,textfont=sf]{subfig}
\else
  \usepackage[caption=false,font=footnotesize]{subfig}
\fi
\usepackage{url}

\makeatletter

\newcommand{\mgSecRef}[1]{Section~\ref{sec:#1}} 
\newcommand{\mgSSecRef}[1]{Section~\ref{ssec:#1}} 
\newcommand{\mgFigRef}[1]{Figure~\ref{fig:#1}} 
\newcommand{\mgTabRef}[1]{Table~\ref{tab:#1}} 
\newcommand{\mgAlgRef}[1]{Algorithm~\ref{alg:#1}} 

\renewcommand{\algorithmicrequire}{\textbf{Input:}}
\renewcommand{\algorithmicensure}{\textbf{Output:}}

\makeatother

\begin{document}

\title{Mining the Workload of Real Grid Computing Systems}

\author{Marco~Guazzone,~\IEEEmembership{Member, IEEE}
\IEEEcompsocitemizethanks{\IEEEcompsocthanksitem M. Guazzone is with the University of Piemonte Orientale, Italy.\protect\\E-mail: \url{marco.guazzone@di.unipmn.com}}}


\IEEEspecialpapernotice{(Technical Report)}


\IEEEtitleabstractindextext{%
\begin{abstract}
Since the mid 1990s, grid computing systems have emerged as an analogy for making computing power as pervasive an easily accessible as an electric power grid.
Since then, grid computing systems have been shown to be able to provide very large amounts of storage and computing power to mainly support the scientific and engineering research on a wide geographic scale.
Understanding the workload characteristics incoming to such systems is a milestone for the design and the tuning of effective resource management strategies.
This is accomplished through the workload characterization, where workload characteristics are analyzed and a possibly realistic model for those is obtained.
In this paper, we study the workload of some real grid systems by using a data mining approach to build a workload model for job interarrival time and runtime, and a Bayesian approach to capture user correlations and usage patterns.
The final model is then validated against the workload coming from a real grid system.
\end{abstract}

}


\maketitle


\section{Introduction} \label{sec:intro}

\emph{Grid computing} \cite{Foster2001Anatomy} is a computing paradigm that has emerged in the mid 1990s as a metaphor to make computing power as pervasively and easily accessible to users as an electric power grid.
Initially conceived to support the research of scientists and engineers, it is now widely adopted also by enterprises \cite{Foster-2004-Grid2}.

According to this paradigm, several collections of heterogeneous and large-scale distributed resources (possibly belonging to different organizations) are loosely federated together to form a single \emph{grid system}, and are shared in a transparent, dynamic and coordinated way to the user community.
The term \emph{virtual organization} (VO) is generally used for referring to as a collection of entities (both users and resources) that belong to multiple independent organizations but that have common goals or shared policies in the context of a particular grid system.

One of the challenging and still open problems in grid computing is the effective management of the resources of a grid system.
In particular, job scheduling is an optimization problem which consists in finding an optimal assignment of computational and storage resources to jobs that are waiting to be executed, in order to minimize the total completion time (i.e., the time taken by every job to execute and to transfer input and output data, if any), without violating any predefined constraint (like site policies).
Since scheduling is an NP-complete problem \cite{Garey1979Computers}, the common way to approach to it is by using an approximated solution through the adoption of a heuristic (i.e., the scheduling strategy), which tries to make good assignments in a reasonable time.
In traditional systems, like cluster systems, schedulers have complete control on local resources and thus can adopt those mechanisms that best adapt to local policies.
For this reason, and for the fact that resources are usually homogeneous, job scheduling in such systems is an affordable task, thus enabling an effective use of local resources.
On the other hand, scheduling in grid systems is a complex task which may involve different, and sometimes conflicting, factors like site autonomy, resource heterogeneity and fault tolerance.
In fact, in grid systems resources are heterogeneous and distributed across different and independent administrative domains, with their own local scheduler usually characterized by different time-varying usage patterns, local policies and security mechanisms.
A grid scheduler should be able to interface and interoperate with these different local schedulers.
Moreover, since resources can be dynamically added and removed, fault-tolerant mechanisms are needed for reacting upon resources unavailability.

For these reasons, a good scheduling strategy, in addition to be efficient and to respect site policies, should take into consideration the workload characteristics of the system where it is applied, in order to be able to adapt to time-varying workload conditions.
Thus, understanding workload characteristics is an essential step for the design and the tuning of effective scheduling strategies.
This is accomplished through the workload characterization, where workload characteristics are analyzed and a possibly realistic model for those is obtained.
Once the workload model is available, it can be used in two possible ways: (1) to generate synthetic workload for simulating or validating scheduling strategies, and (2) to create adaptive scheduling algorithms for predicting future workload.

In this paper, we model workload characteristics of some real grid systems by using a data mining approach.
In particular, we use cluster analysis as a primary tool for compactly representing workload characteristics.
The result of cluster analysis was then employed to build a model of the job interarrival time and runtime workload characteristics.
We focus on these two characteristics since they are among the most important from the point of view of a grid scheduler.
Besides the use of clustering analysis, we employ a Bayesian approach to model correlations between users and job submissions in order to capture user behavioral patterns.
The final model in then validated against the workload coming from a real grid system.

The rest of this paper is structured as follows.
In \mgSecRef{method}, we describe the methodology we use for analyzing the workload characteristics and building a workload model.
In \mgSecRef{exp}, we present the analysis and the modeling of workload coming from a real grid system.
Finally, in \mgSecRef{conclusion}, we conclude this paper and we present possible future works.


%
%
%


\section{Methodology} \label{sec:method}

\subsection{Modeling Approach} \label{ssec:method-approach}

In this section, we present the approach we use for analyzing and modeling the data.
Our approach follows the classical \emph{Knowledge Discovery from Data} (\emph{KDD} process) \cite{Shapiro1991Knowledge,Fayyad1996DataOverview}, a process which seeks new and usually hidden knowledge (about a specific application domain) from data, and which consists of multiple sequential steps: \emph{understanding the application domain}, \emph{data collection}, \emph{data selection}, \emph{data cleaning and preprocessing}, \emph{data transformation}, \emph{data mining}, \emph{model assessment and validation}, \emph{knowledge consolidation}.
KDD is an iterative process, with many feedback loops and repetitions, which are triggered by revision phases.

KDD starts with the \emph{understanding the application domain} step, which consists in understanding the problem domain and the relevant prior knowledge in order to identify the KDD goals from the point of view of the domain user.

Then, KDD continues with the \emph{data collection} step, which involves collecting the data that will successively be analyzed; this phase consists, firstly, in defining what attributes provide enough information for the subsequent steps, and, secondly, in setting up the proper instrumentation tools for collecting the data.

After data have been collected, KDD proceeds with the \emph{data selection} step where a portion of the whole data is sampled; data should be extracted in a way that the size of the resulting set would be large enough to contain meaningful information but small enough to manipulate it quickly.

Subsequently, the selected data might need to be processed in order to remove or fix possible anomalies.
This is the objective of the \emph{data cleaning and preprocessing} step, which consists in looking at the data in order to find out anomalous data like missing values, errors, unexpected values and, generally, any kind of suspicious values.
This step has to be carried out with care because even the removal of few values can alter the result of future analysis; usually, when there is little or no knowledge about the collected data it is advisable to avoid to remove or change data, unless the anomaly is evident.

Once the selected data have been preprocessed and before moving to the pattern analysis, it might be useful to transform them for finding useful attributes and invariant representation.
This is done in the \emph{data transformation} step, by applying, for instance, dimensionality reduction, normalization and other transformation methods.
This step is especially useful when there are several attributes (i.e., high-order data) or when two or more attributes, with different order of magnitude, are involved in the same operation (e.g., distance or similarity calculation).

After the selected data have been filtered and transformed, it is possible to try to match the goals defined in the first step.
This is done in the \emph{data mining} step, where data are explored, in order to understand their underlying structure and potential relationships between the different attributes, and a model for them is built.
It consists of some iterative subordinate phases in which the analyst has to choose a particular data mining method (e.g., summarization, classification, regression, clustering, and so on) for matching the predetermined goals, to select a suitable algorithm through which carry on the chosen data mining method, and finally to search for interesting patterns and evaluate their significance.
The data mining phase outputs a model which provides a global or local description of the analyzed data by generating patterns in a particular representational form, such as classification rules, decision trees or regression models, according to the selected data mining algorithm.

Once a model for the data has been built, it is important to evaluate its performance and accuracy for determining how much good it fits to the data, how much good it generalizes on unseen observations and, in general, how much it is representative of the knowledge with respect to a specific measure of interestingness. 
This is accomplished in the \emph{model assessment and validation} step.
This step can also include comparison with other preexisting models, for comparing their performance and choosing the one that is the more statistically accurate.
Depending on how good is the built model, this phase may require to return to any of the previous steps for further iterations and refinements.

Finally, the last step, the \emph{knowledge consolidation} step, consists in using the discovered knowledge and possibly incorporating it into an existing system.

\subsection{Workload Analysis} \label{ssec:method-analysis}

In this work, we follow the aforementioned KDD process in a way as strictly as possible, but with some exceptions.
Firstly, we skip the data collection and selection phases (the second and third step of KDD, respectively) since the data we use have been collected by third party entities.
Moreover, because of the lack of information about the data collection phase, we have to proceed with extremely care in subsequent phases, especially during the data cleaning and preprocessing step (the fourth step of KDD), in order to avoid the removal of important and meaningful information.
Specifically, in data cleaning and preprocessing step, we opt to remove or to fix only those data that show evident anomalies or those that don't change the underlying marginal probability distribution, leaving the other possible candidates intact; the reasons about this choice are to be mainly ascribed to the possible bursty behavior of the system (where few large values strongly influence the other ones and the average case), which causes the average case to be not always a good representative of the real system behavior.
Finally, in the data mining phase (the fifth step of KDD), before creating the workload model, we investigate for several statistical properties of both the marginal and the joint probability distributions of the interested attributes which may strong influence the selection of the modeling tool and the quantitative analysis of the system behavior.
Such statistical properties include:
\begin{itemize}
\item \emph{normality} \cite{Montgomery2002Applied} (whereby a probability distribution can be accurately approximated by a normal distribution),
\item \emph{long-tailness} and \emph{power-law} \cite{Newman2005PowerLaws} (where regions far from the mean or the median of the probability distribution, like the extreme values in the tails of the probability distribution, are assigned relatively high probabilities following a sub-exponential or polynomial law, contrary to what happens to the family of Normal distributions, where the tails fall exponentially),
\item \emph{autocorrelation}, including \emph{short-range dependence} \cite{Brockwell2002Introduction} (also known as serial correlation or short-term memory, where the autocorrelation at short time scales is significant) and \emph{long-range dependence} \cite{Beran1994Statistics} (also known as long-term memory, where the autocorrelation at long time scales is significant),
\item \emph{self-similarity} \cite{Embrechts2000Introduction} (where, in addition to the long-range dependence property, the scale invariance property holds at any different time scale, like in a fractal shape),
\item \emph{heavy-tailness} \cite{Adler1998Practical} (where the long tails of the distribution fall polynomially and the self-similarity property holds),
\item \emph{cyclic behavior} and \emph{seasonal variations} \cite{Brockwell2002Introduction} (which are an indication of a non-stationary workload and must be treated separately).
\end{itemize}
In order to carry on this investigation, we use both graphical and quantitative tools coming from statistical data analysis.
For the graphical evaluation, we follow the \emph{exploratory data analysis} (EDA) approach \cite{Tukey1977EDA}, by using:
\begin{itemize}
\item The \emph{log-log EDF} and \emph{log-log CEDF plots}, for comparing the body and the tails of the sample probability distribution, respectively.
This is achieved by plotting the \emph{empirical cumulative distribution function} (EDF) and the \emph{complementary EDF} (CEDF), respectively, of two probability distributions of interest; if the plots are nearly overlapping, it means that the two probability distributions are statistically equivalent.
\item The \emph{Q-Q plot} \cite{Wilk1968Probability}, for evaluating possible differences, especially in the tails, between the empirical probability distribution and another reference probability distribution, either theoretical or empirical.
This is achieved by plotting the quantiles of the two probability distributions of interest against each other; if the points of the plot approximatively lie on the \emph{reference line} (i.e., on the line $y=x$) it means that the two probability distributions are statistically similar.
\item The \emph{mass-count disparity plot} \cite{Crovella2001Performance,Feitelson2006Metrics} and the \emph{Lorenz curve} \cite{Lorenz1905Methods}, for looking for an evidence of the power-law property.
The mass-count disparity plot consists in plotting the ``mass'' probability distribution (given by the probability that a unit of mass belong to an item smaller than a specified $x$) against the ``count'' probability distribution (given by the cumulative distribution function) in order to show possible disparities between these two probability distributions.
The rational here is that, in the presence of a power-law distribution, the disparity between the two distribution should be very large since a small number of samples account for the majority of mass, whereas all small samples together only account for negligible mass (e.g., if the job runtime distribution was a power-law distribution, the majority of jobs would have short runtimes and only few jobs would have very long runtimes; however, these few jobs would contribute to the majority of the mass of the runtime distribution).
The analysis of this plot consists in evaluating the area between the two curves; the larger is this area, the higher is the likelihood that the distribution of interest is a power-law distribution.
The Lorenz curve is another way to display the relationship between the ``count'' distribution and the ``mass'' distribution; it is constructed by pairing percentiles that correspond to the same value (i.e., a point $(p_c,p_m)$ in the curve is such that $p_m=F_m(x)=F_m(F_c^{-1}(p_c))$ where $F_m(\cdot)$ and $F_c(\cdot)$ are the cumulative distribution functions of the ``mass'' and ``count'' distribution, respectively, and $F_c^{-1}(\cdot)$ is the inverse of $F_c(\cdot)$).
The analysis of this plot consists in looking at the distance of the curve from the \emph{line of perfect equality} (i.e., the $y=x$ line); the larger is the distance the higher is the chance that the distribution of interest is a power-law distribution; such distance can also be evaluated by computing the \emph{Gini coefficient}.
\item The \emph{run-sequence plot} \cite{Chambers1983Graphical} and the \emph{autocorrelation plot} \cite{Box1970Time} for investigating for the presence of both short-range and long-range dependence as long as for periodic patterns and trends.
The run-sequence plot displays observed data in a time sequence; it is constructed by plotting values of the interested (univariate) attribute according to the temporal order as they appear; this plot is particularly useful for finding both shifts in location and scale, for locating outliers and, in general, for getting insights about the trend of observed data. 
The autocorrelation plot (also known as \emph{correlogram}) is a plot of the sample autocorrelation function (ACF), that is of the sample autocorrelation at increasing time lags;
it is used for checking for randomness in a sequence of observations of the same attribute.
If random, such autocorrelations should be near to $0$ for any and all time-lag separations; conversely, if non-random, then one or more of the autocorrelations will be significantly different from $0$. 
\item The \emph{scatter-matrix plot} \cite{Chambers1983Graphical} for looking for cross-correlation between different attributes of the same data set; this plot consists in drawing a scatter plot for each possible pair of attributes;, on each scatter-plot, the value of an attribute is plotted against the corresponding value (i.e., the one found in the same observation of the data set) of another attribute.
\end{itemize}
For what concerns the quantitative evaluation, we use classical tools of the descriptive statistics like the following:
\begin{itemize}
\item The \emph{two-sample pooled (Welch) $t$-test}, for comparing, in a parametric way, the mean of two distributions (with possibly different variances); it assumes that the two samples under study come from a (possible asymptotically) Normal distribution.
\item The \emph{$F$-test}, for comparing, in a parametric way, the variance of two distributions; it assumes that the two samples under study come from a (possible asymptotically) Normal distribution.
\item The \emph{Mann-Whitney $U$-test} (also known as the \emph{Wilcoxon rank-sum test}), for assessing, in a non-parametric way, whether two independent samples come from the same distribution against the alternative hypothesis that the two distributions differ only with respect to the median; it assumes that within each sample under study the observations are independent and identically distributed, and that the shapes and spreads of the distributions are the same.
\item The \emph{Ansari-Bradley test}, for assessing, in a non-parametric way, whether two independent samples come from the same distribution against the alternative hypothesis that the two distributions differ only in scale; it assumes that within each sample under study the observations are independent and identically distributed, and that the two samples must
be independent of each other, with equal medians.
\item The \emph{two-sample Kolmogorov-Smirnov $K$-$S$ test}, for comparing two empirical distribution functions.
\item The \emph{Pearson's $r$} and the \emph{Spearman's $\rho$ correlation coefficients}, for discovering linear and generic correlations, respectively, among the interested attributes; both coefficients take values in the interval $[-1,+1]$, where $-1$ means a strong negative correlation (i.e., if the value of one attribute increases, the value of the other attribute decreases, and vice versa), $+1$ means a strong positive correlation (i.e., if the value of one attribute increases, the value of the other attribute increases too, and vice versa), and a value of $0$ means no significant correlation; for all the other values in this interval, the nearest are to $0$, the weaker is the correlation.
\end{itemize}
Some of these tools have also been used in the data cleaning and processing step for evaluating any difference between the original marginal distribution of the attribute of interests and the one obtained after removing those values considered anomalous.
Specifically, we use graphical tools like the log-log EDF plot, the log-log CEDF plot and the Q-Q plot for qualitatively comparing the empirical distribution function, the complementary empirical distribution function and the quantile function of the two distributions, respectively, and numerical tools like the two-sample Kolmogorov-Smirnov $K$-$S$ test, for quantitatively assessing any difference in the empirical distribution function of the two sample distributions, and the two-sample pooled $t$-test, the Mann-Whitney $U$-test, the $F$-test and the Ansari-Bradley test, for numerically evaluating any difference in the location and in the scale of the two sample distributions.

\subsection{Workload Model Construction} \label{ssec:method-model}

For modeling workload characteristics we need a modeling framework that is able to preserve both first-order and second-order statistics (e.g., quantiles and correlation, respectively) present in the data.
Traditional statistical univariate distribution fitting lacks of such ability since it enables only to model marginal probability distributions and hence first-order statistics.
However, in real workloads, it is often the case that some of the workload characteristics are tight together according to some kind of correlation.
While correlation can be easily discovered through some graphical or numerical tools, like the ones introduced above, it is difficult to model with only statistical distributions since one would have to consider multivariate statistical distributions, something that is hard to fit.
In order to model such correlations, we apply \emph{cluster analysis} \cite{Tan2006Introduction} to summarize workload characteristics, in order to identify representative values for each workload characteristic, and a Bayesian approach in order to exploit the locality of sampling principle \cite{Feitelson2007Locality}.
Specifically, we apply cluster analysis to summarize job submissions in terms of $\langle\text{\texttt{iatime}},\text{\texttt{runtime}}\rangle$ pairs, and we use a Bayesian approach to estimate the probability distribution of user job submissions.

Cluster analysis divides data into groups (clusters) according to some similarity concept.
The purpose of cluster analysis is two-fold: (1) it allows to capture the inherent structure of the data (when it exists), and (2) it provides a way to summarize data.
For these reasons, cluster analysis is an appealing way to characterize data.
In literature, there can be find several type of cluster algorithms.
In this work, we consider only \emph{partitional algorithms}, that is algorithms that simply divide the set of data into non-overlapping subsets (clusters) such that each data point is in exactly one subset.
Such algorithms can be further classified in several sub-categories, according to how they partition the data set.
Specifically, in this work we employ three different partitional algorithms:
\begin{itemize}
\item \emph{CLARA} \cite{Kaufman1990Finding}, a prototype-based cluster algorithm suitable for large data set;
\item \emph{MCLUST} \cite{Fraley1999MCLUST}, a (probabilistic) model-based cluster algorithm based on multivariate Gaussian fitting;
\item \emph{DBSCAN} \cite{Ester1996Density}, a density-based cluster algorithm suitable for large data set.
\end{itemize}
This choice has been primarily influenced by the large size of the data set we consider.
We now briefly illustrate how such algorithms work.

The \emph{Clustering LARge Applications} (\emph{CLARA}) algorithm is a prototype-based cluster algorithm which derives from the \emph{Partitional Around Medoids} (\emph{PAM}) \cite{Kaufman1990Finding} algorithm.
The PAM algorithm is a $k$-medoids cluster algorithm which attempts to cluster a set of $n$ points into $k$ partitions (clusters) so that the sum of the dissimilarities of all the objects to their nearest representative object (medoid) is minimized.
After an initial random selection of $k$ objects as medoids, PAM generates $k$ clusters by repeatedly trying to select the best $k$ representatives objects (medoids) according to a cost function based on object dissimilarities.
Once the $k$ medoids are found, the $k$ clusters are constructed by assigning each object to the nearest medoid.
The most important weakness of PAM is its computational complexity (i.e., $\mathcal{O}(n^2)$ for both time and space, where $n$ is the number of observations), which does not allow PAM to scale well for large data sets.
The CLARA algorithm is an adaptation of PAM for handling large data sets.
It works by repeatedly sampling a set of data points and by applying to each sample the PAM algorithm in order to find the best $k$ ``sample'' medoids.
At last, the set of $k$ ``sample'' medoids that minimizes a cost function is selected.
The computational complexity of CLARA is $\mathcal{O}(ks^2+k(n-k))$, where $s$ is the sample size, $k$ is the number of clusters, and $n$ is the total number of objects.
CLARA takes as input parameters the number of cluster to look for, the sample size and the number of samplings.

The other algorithm we use is the \emph{Multivariate Normal Mixture Modeling and Model-Based Clustering} (MCLUST) algorithm, a probabilistic model-based cluster algorithm where data is considered to be a sample independently drawn from a multivariate Gaussian mixture model, and each cluster is represented by a different multivariate Gaussian distribution belonging to the mixture model.
In MCLUST, the problem of finding $k$ clusters is reduced to the problem of estimating the parameters of the multivariate Gaussian mixture model from data; this is accomplished by combining agglomerative hierarchical clustering and the \emph{Expectation-Maximization} (EM) algorithm \cite{Dempster1977Maximum}.
The problem for determining the number of clusters is reduced to a model selection problem and is approached by using a Bayesian selection model based on the \emph{Bayesian Information Criterion} (BIC) \cite{Schwarz1978Estimating}.
MCLUST combines these aspects by performing the following steps: (1) determines a maximum number $k$ of clusters, (2) performs agglomerative hierarchical clustering and obtain the corresponding classification for up to $k$ groups, (3) applies the EM algorithm for each parameterization and each number of clusters from $2$ to $k$, (4) computes the BIC value for the one-cluster case for each model and for the mixture model with the optimal parameters from EM for all the clusters from $2$ to $k$, and (5) selects the model which gives the best BIC value, which in turns gives strong evidence for a fitted model.
MCLUST takes as input parameter only the maximum number of clusters to look for.

Finally, the remaining algorithm we use is the \emph{Density-Based Spatial Clustering of Applications with Noise} (DBSCAN) algorithm, a density-based cluster algorithm which is able to find clusters of arbitrary shape (even in presence of noise).
DBSCAN looks for clusters by checking the $\epsilon$-neighborhood (i.e., the neighborhood within a radius of $\epsilon$) of each point in the data set.
If the $\epsilon$-neighborhood of a point $p$ contains more than \emph{MinPts} points, a new cluster with $p$ as a core (i.e., centroid) object is created.
DBSCAN then iteratively aggregates neighbor objects with respect to these core points, possibly merging near clusters.
The process terminates when no new point can be added to any cluster.
At last, the resulting clusters are represented by maximal sets of density-connected points, that is by the largest sets of points each of which contains points which are reachable by a chain of neighbor points that are far from each other at most by $\epsilon$.
If no special data structure is used, the computational complexity of DBSCAN is $\mathcal{O}(n^2)$.
DBSCAN takes as input parameters the radius $\epsilon$ and the minimum number of points \emph{MinPts}.

It is important to note that cluster analysis is not a ``panacea'' and thus it should be applied with care.
As a matter of fact, a key motivation is that almost every clustering algorithm will find clusters in a data set, whether the data are naturally clustered or purely random; this happens since any clustering algorithm will impose a clustering structure due to inherent properties of the chosen clustering algorithm and to parameters passed to it (e.g., with the $k$-means algorithm, $k$ clusters are always found).
Moreover, even if data exhibits natural clusters, different clustering algorithms generally find different clusters; thus, after applying a clustering algorithm, ones should verify that the obtained result is really meaningful.
For these reasons, clustering analysis should always include two parts: (1) \emph{clustering tendency}, and (2) \emph{clustering validation} (or \emph{clustering evaluation}).

\emph{Clustering tendency} concerns with determining, before applying any clustering algorithm, if a data set is naturally clustered, in order to distinguish whether a some kind of structure, other than random, actually exists in the data.
For clustering tendency, several methods have been developed that check for the existence of subgroups with higher homogeneity; in particular, they divide in graphical and numerical methods.

Graphical methods consist in graphically representing data in order to highlight possible clusters.
Actually, most of them are based on the concept of \emph{heat-map} \cite{Ling1973Computer,Bezdek2002VAT}, a false-color image with an optional \emph{dendrogram} (i.e., a tree-like visualization of hierarchical clusters) added to left side and/or to the top, that simultaneously reveals row and column hierarchical cluster structure in a data matrix.
Essentially, a heat-map consists of a rectangular tiling with each tile shaded on a color scale to represent the value of the corresponding element of the data matrix; the rows (columns) of the data matrix are re-ordered such that similar rows (columns) are near each other.
Historically, heat-maps are used for unsupervised clustering validation, specifically, as a tool for validating the result of hierarchical clustering.
However, recently it has been proposed a new method for creating heat-maps for unsupervised clustering tendency, named \emph{Visual Assessment Tendency} (\emph{VAT}) \cite{Bezdek2002VAT}; this method consists in creating a heat-map by reordering the dissimilarity matrix according to a minimum-spanning tree like criterion.
One disadvantage of visual methods based on the notion of heat-map, lies in the fact that they work directly with similarity (or dissimilarity) matrices; this means that such methods are not suitable for very large data sets because of too large memory requirements.
A possible workaround to the dimensionality problem is to compute the similarity matrix on only a small sample of the input data and hence to create the heat-map on that matrix; however, it is important to note that since the sample might not be representative of the data (i.e., might not capture the natural structure of the whole data set) this process should be repeated several times (through sequential resamplings) in order to increase the probability to obtain a significant sample (e.g., see \cite{Tan2006Introduction}).
Indeed, we use this workaround to produce VAT images in a way similar to the one proposed by \cite{Park2009Visualisation}.

For what concerns numerical methods for clustering tendency, all of them are essentially stated in terms of \emph{internal criterion} (i.e., no additional information is used other than data itself) and can be viewed as the problem of testing for spatial randomness (i.e., data points are uniformly distributed in the data space) or, similarly, of fitting a spatial point process to data \cite{Jain1988Algorithms}.
A test for clustering tendency can be thought as a statistical hypothesis test where the null hypothesis $\mathcal{H}_0$ is ``no structure in data'', while the alternative hypothesis $\mathcal{H}_1$ is ``data has natural clusters''.
If this null hypothesis cannot be rejected, the result of any cluster analysis procedure will be only a random partitioning of the objects, depending on the actual algorithm used.
There are two key points that have an important influence on the performance of many statistical tests that are used in clustering tendency.
The first point is the dimensionality of data which has a great impact on the performance.
The other key point is the size of the so called \emph{sampling window}; the sampling window can be viewed as a set $S$ in a $d$-dimensional space (where $d$ is the dimension of the data space, that is the number of attributes) used for testing (under the null hypothesis) that data are uniformly distributed over $S$ and hence that they have no natural clusters.
The choice of the size of the sampling window is critical since the same data can appear as random or non-random by only varying the sampling-window size.
Among the several numerical methods, we use the \emph{Hopkins test} \cite{Hopkins1954New} since it has been claimed to be the more powerful one \cite{Zeng1985Comparison}.
The Hopkins statistic compares Euclidean distances of sample data objects to the related nearest-neighboring objects, with Euclidean distances of arbitrary artificial points (uniformly generated in the data space) to the related nearest-neighboring objects.
Specifically, a number $m$ of artificial points are randomly generated by uniformly sampling the $d$-dimensional data space; further, a number $m$ of data points are randomly selected.
For both sets of points, distances to the nearest neighbors in the original data set are computed.
Let $u_i$ be the nearest-neighbor distance of the $i$-th artificially generated point, and $w_i$ be the nearest-neighbor distance of the $i$-th sampled point.
The Hopkins statistics $H$ is defined as:
\begin{equation}
\begin{aligned}
H &= \frac{\sum_{i=1}^m{u_i^d}}{\sum_{i=1}^m{u_i^d}+\sum_{i=1}^m{w_i^d}} \\
  &= \left(\frac{\sum_{i=1}^m{w_i^d}}{\sum_{i=1}^m{u_i^d}}+1\right)^{-1}
\end{aligned}
\end{equation}
From the above equations (especially, from the second one), it is clear that, in presence of clustering tendency, $w_i$ will tend to be smaller than $u_i$ and thus $H$ will be larger than $0.5$ and at most be $1$.
Practically, the Hopkins statistics is computed for several random selection of points, and the average of all results for $H$ is used for a decision; if this average is greater than $0.75$ then the null hypothesis can be rejected with high (i.e., $90\%$) confidence.
The value of $H$ depends on the choice of $m$, which is the size of the sampling window; usually $m$ is taken to be equal to the $10\%$ of the size of the entire data set (i.e., the number of observations).
In literature several modifications to this test have been proposed for mitigating the sampling window problem (e.g., see \cite{Fernandez2000Improved}).
It is worth noting that the Hopkins test, as well as the other numerical test, comes with several issues.
Firstly, as already discussed, the choice of the size of sampling window plays a central role in the numerical result of the method.
Secondly, numerical methods are usually defined in terms of Euclidean distance; however, for specific data set, other distance metrics are more suitable.
Finally, they are computationally inefficient when the sizes of the sampling window and of the data set are large.

The other important part of cluster analysis is \emph{clustering validation}, which consists in verifying, after having applied any clustering algorithm, if the result obtained from the clustering algorithm is significant.
It includes the followings task:
\begin{itemize}
\item Determining the right number of clusters.
\item Evaluating the accuracy of the obtained clustering model against internal information.
\item Evaluating the accuracy of the obtained clustering model against external information.
\end{itemize}
In literature, there exist several graphical and numerical tools for assessing clustering validation; essentially, they divide in \emph{internal} (or \emph{unsupervised}) \emph{criteria}, \emph{external} (or \emph{supervised}) \emph{criteria} and \emph{relative criteria}.
Internal criteria do not use external information; they take only the data set and the clustering partition as input and use intrinsic information in the data to assess the quality of the clustering.
External criteria use external information; they measure the extent to which the clustering structure discovered by a clustering algorithm matches a given external structure.
Relative criteria compare the result of different clustering algorithms or of the same algorithms but with different parameters (e.g., the number of clusters).
For prototype-based cluster algorithms, one of the most popular method for assessing clustering validity is the \emph{silhouette coefficient} \cite{Rousseeuw1987Silhouettes}, an internal criterion which measure the quality of a given cluster; given a point $i$, the silhouette width for $i$ is defined as:
\begin{equation}
s_i = \frac{b_i-a_i}{\max\{a_i,b_i\}}
\end{equation}
where $a_i$ is the average distance of $i$ to all other points in the same cluster, and $b_i$ is the minimum between the average distances of $i$ to all the points in any other clusters (but the one containing $i$).
The silhouette coefficient $SC$ is thus obtained by averaging over all silhouette widths.
The silhouette coefficient is always between $-1$ and $1$.
For the purpose of validation, a common interpretation of $SC$ values is the following:
\begin{itemize}
\item If $SC \in (0.70, 1.00]$, the clustering structure is strong.
\item If $SC \in (0.50, 0.70]$, the clustering structure is plausible.
\item If $SC \in (0.25, 0.50]$, the clustering structure is weak.
\item If $SC \in [0.00, 0.25]$, lack of clustering structure.
\end{itemize}
For model-based cluster algorithms, each algorithm usually embeds an optimal model selection criterion which is used for determining both model parameters and the number of clusters; the most widely used criterion is the \emph{Bayesian Information Criterion} (BIC) \cite{Schwarz1978Estimating}, which is defined as:
\begin{equation}
\text{BIC}_k = 2\log(\Pr\{D|\hat{\theta}_k,M_k\})-\nu_k\log(n)
\end{equation}
where $D$ is the observed data, $M_k$ is a specific model, $\hat{\theta}_k$ are the estimated parameters of model $M_k$, $n$ is the number of data points in $D$ (i.e., the number of observations), and $\nu_k$ is the number of free parameters to be estimated in model $M_k$.
Given any two estimated models, the model with the lower value of BIC is the one to be preferred. 
For what concerns density-based cluster algorithms, to our knowledge, there is no well-proven clustering validation technique.

After applying cluster analysis and choosing the best result (according to the above clustering validation measures), we characterize user submissions in a probabilistic way (by using a Bayesian approach) in order to maintain possible locality structures.
Specifically, we create a workload model $\mathcal{M}=\langle\mathcal{C},\mathbf{u},\mathcal{W}, \mathbf{S} \rangle$ according to the following steps:
\begin{algorithm*}
\caption{The workload model construction algorithm}\label{alg:workload-model}
\begin{algorithmic}[1]
\Function{CreateWorkloadModel}{}
	\State Create and normalize the input data set in order to contain only pairs of interarrival times and runtimes.
	\State Apply cluster analysis to interarrival and runtime attributes for obtaining the set of clusters $\mathcal{C}=\{C_1,\ldots,C_m\}$, such that $C_i$ is a cluster of centroid $(a_i,r_i)$ where $a_i$ and $r_i$ are the representatives of interarrival times and runtimes inside cluster $C_i$, respectively.
	\State Create a vector $\mathbf{u}=(u_1,\ldots,u_n)$ representing the fraction of job submissions given by each user; that is, given a user $i$, the $i$-th component $u_i$ of vector $\mathbf{u}$ is defined as:
	\begin{equation}
	u_i = \frac{\text{\# jobs submitted by user $i$}}{\text{\# jobs submitted by all users}},\quad i=1,\ldots,n
	\end{equation}
	obviously it results that $\sum_{i=1}^n u_i=1$.
	The vector $\mathbf{u}$ can be viewed as a discrete probability mass function giving the probability $p_u(\cdot)$ that a specific user submits a job; specifically, the probability $p_u(i)$ that user $i$ submits a job is given by:
	\begin{equation}
	p_u(i)=u_i,\quad i=1,\ldots,n
	\end{equation}
	\State Create the set of vectors $\mathcal{W}=\{\mathbf{w}_1,\ldots,\mathbf{w}_n\}$ such that each vector $\mathbf{w}_i=(w_{i1},\ldots,w_{im})$ represents the fraction of workload contributed by user $i$ for each cluster:
	\begin{equation}
	w_{ij} = \frac{\text{\# jobs in cluster $C_j$ belonging to user $i$}}{\text{\# jobs submitted by user $i$}},\quad i=1,\ldots,n \text{ and } j=\,\ldots,m
	\end{equation}
	where $\sum_{j=1}^m w_{ij}=1$, for each $i=1,\ldots,n$.
	The vector $\mathbf{w}$ can be viewed as a discrete conditional probability mass function giving the probability $p_w(\cdot|i)$ that a specific user $i$ submits a job characterized by the interarrival time and the runtime summarized by a particular cluster, given that user $i$ submits a job; specifically, the probability $p_w(j|i)$ that user $i$ submits a job with interarrival time and runtime represented by the centroid of cluster $C_j$, given that user $i$ submits a job, is defined as:
	\begin{equation}
	p_w(j|i)=w_{ij},\quad i=1,\ldots,n \text{ and } j=1,\ldots,m
	\end{equation}
	Together, the quantities $p_u(i)$ and $p_w(j|i)$ can be viewed as the probability that a user $i$ submit a job characterized by cluster $C_j$, and can be thought as a Bayesian way to characterize user job submissions.
	\State Denormalize the input data set.
	\State Create an $m \times 2$ matrix $\mathbf{S}=(s_{ij})$, where each row $i$ contains the unbiased sample standard deviations of interarrival time (first column) and runtime (second column) inside cluster $C_i$.
	This matrix can be used during synthetic workload generation in order to estimate data variability.

	\Return $\mathcal{M}=\langle \mathcal{C},\mathbf{u},\mathcal{W},\mathbf{S} \rangle$
\EndFunction
\end{algorithmic}
\end{algorithm*}

\subsection{Synthetic Workload Generation}

The workload model $\mathcal{M}=\langle \mathcal{C},\mathbf{u},\mathcal{W},\mathbf{S} \rangle$ obtained by means of \mgAlgRef{workload-model}, described in the previous section, can then be used to generate synthetic workload traces.
This can be accomplished by means of \mgAlgRef{synthetic-workload}.

The rationale underlying this algorithm is that correlations between workload characteristics (obtained through cluster analysis) have to be weighted by the actual contribution that each user provides to the workload.
The workload contribution of each user $i$ is modeled through vectors $\mathbf{u}$ and $\mathbf{w}_i$ which together provide a Bayesian way to compute the probability $u_i*w_{ij}$ that a specific user $i$ submits a job represented by a particular cluster $C_j$.
\begin{algorithm*}
\caption{The synthetic workload generation algorithm}\label{alg:synthetic-workload}
\begin{algorithmic}[1]
\Function{GenerateSynthWorkload}{$\mathcal{M}=\langle \mathcal{C},\mathbf{u},\mathcal{W},\mathbf{S} \rangle, N$}
	\Require $\mathcal{C}=\{C_1,\ldots,C_m\}$ is the set of clusters. 
	\Require $\mathbf{u}=(u_1,\ldots,u_n)$ is the vector of user weights. 
	\Require $\mathcal{W}=\{\mathbf{w}_1,\ldots,\mathbf{w}_n\}$ is the set of vector of cluster weights for each user. 
	\Require $\mathbf{S}=(s_{ij})$ a $m \times 2$ matrix containing the standard errors of attributes for each cluster. 
	\Require $N$ is the size of the final synthetic trace. 
	\Ensure A set $T$ of size $N$. 
	\State $T \gets \emptyset$ \Comment{The final synthetic workload trace.}
	\Repeat
		\State $r_u \gets \Call{UnifRand}{0,1}$ \Comment{Randomly generate a number uniformly distributed in $[0,1]$}
		\State $i^* \gets \max \{i| \sum_{k=1}^i u_k \le r_u, i=1,\ldots,n\}$
		\State $r_w \gets \Call{UnifRand}{0,1}$ \Comment{Randomly generate a number uniformly distributed in $[0,1]$}
		\State $j^* \gets \max \{j| \sum_{k=1}^j w_{i^*k} \le r_w, j=1,\ldots,m\}$
		\State $(\hat{a}_j, \hat{r}_j) \gets \Call{Centroid}{C_{j^*}}$
		\State $a \gets \Call{ExpRand}{\hat{a}_j}$ \Comment{Sample from an Exponential distribution with mean $\hat{a}_j$}
		\State $r \gets \Call{ExpRand}{\hat{r}_j}$ \Comment{Sample from an Exponential distribution with mean $\hat{r}_j$}
		\State $T \gets T \cup \{(a,r)\}$
	\Until $\lvert T \rvert = N$

	\Return $T$
\EndFunction
\end{algorithmic}
\end{algorithm*}




\section{Experimental Evaluation} \label{sec:exp}

In this section, we present the characterization of the workload coming from the \emph{Large Hadron Collider Computing grid} (\emph{LCG}) system \cite{Cern_LCG_Url}.
The LCG trace contains data corresponding to $11$ days of activity (from November $20^\text{th}$ to $30^\text{th}$ in $2005$) from multiple machines that constitute the LCG system.
This log was previously analyzed and described in \cite{Li2006Modeling} and has been graciously provided by the \emph{Parallel Workload Archive} (PWA) \cite{Feitelson_PWA_Url}.
As described in that work, data are collected at the level of grid resource brokers, and hence does not include any locally generated load.
The total number of entries is $188041$.
Each entry in the log represents a possibly parallel job.
Moreover, the log is at the level of individual processes, and does not contain data about which processes may be part of the same parallel job.
As a consequence there is no information about job parallelism and hence all jobs (actually processes) are recorded as having a size of $1$.
The log format includes five attributes:
\begin{itemize}
\item \texttt{timestamp}: a numerical integer attribute representing the timestamp of the job submission time, expressed as Unix epochs, that is as the number of seconds elapsed since $00$:$00$:$00$ on January $1$, $1970$ Coordinated Universal Time (UTC).
\item \texttt{uid}: a categorical attribute representing the identifier of the user who submitted the job.
\item \texttt{vo}: a categorical attribute representing the name of the Virtual Organization to which the submitting user belongs.
\item \texttt{ce}: a categorical attribute representing the name of the computing element where the job has executed.
\item \texttt{runtime}: a numerical value representing the duration of the job execution, expressed in seconds.
\end{itemize}
It is important to note that even if the attribute \texttt{uid} has integer values, it is to be considered as a categorical attribute because it is not possible to define an order between its values (e.g., user identifier $401$ is neither less than nor greater to user identifier $10$).
Since we are interested in studying the behaviour of the running time as well as the interarrival time, we extend the original log format with the derived attribute \texttt{iatime}, a numerical integer attribute representing the job interarrival time, and computed it by subtracting to each submission (arrival) time the immediately preceding arrival time, that is:
\begin{align*}
\text{\texttt{iatime}}_1 &= 0\\
\text{\texttt{iatime}}_i &= \text{\texttt{timestamp}}_{i} - \text{\texttt{timestamp}}_{i-1}, \quad i>1
\end{align*}

To perform the characterization of the LCG workload, we develop a set of statistical libraries in \texttt{R}, that are publicly available for the sake of research reproducibility \cite{dcsr}.
All the experiments presented in this section are carried out on a dedicated Intel Quad-Core Xeon $X3220$ machine equipped with $4$GB of RAM and running the $64$ bit Linux $2.6.30$ operating system.

\subsection{Data Cleaning and Preprocessing} \label{ssec:exp-filter}

\begin{figure*}
\centering
\subfloat[Comparison of EDFs (in log-scale).]{\includegraphics[scale=0.17]{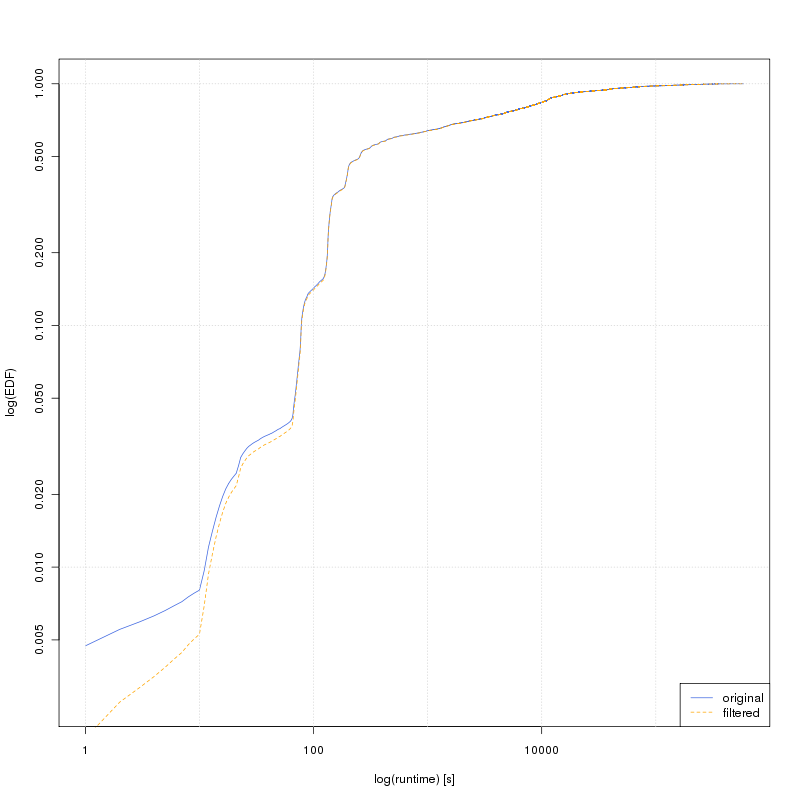}}
\quad
\subfloat[Comparison of CEDFs (in log-scale).]{\includegraphics[scale=0.17]{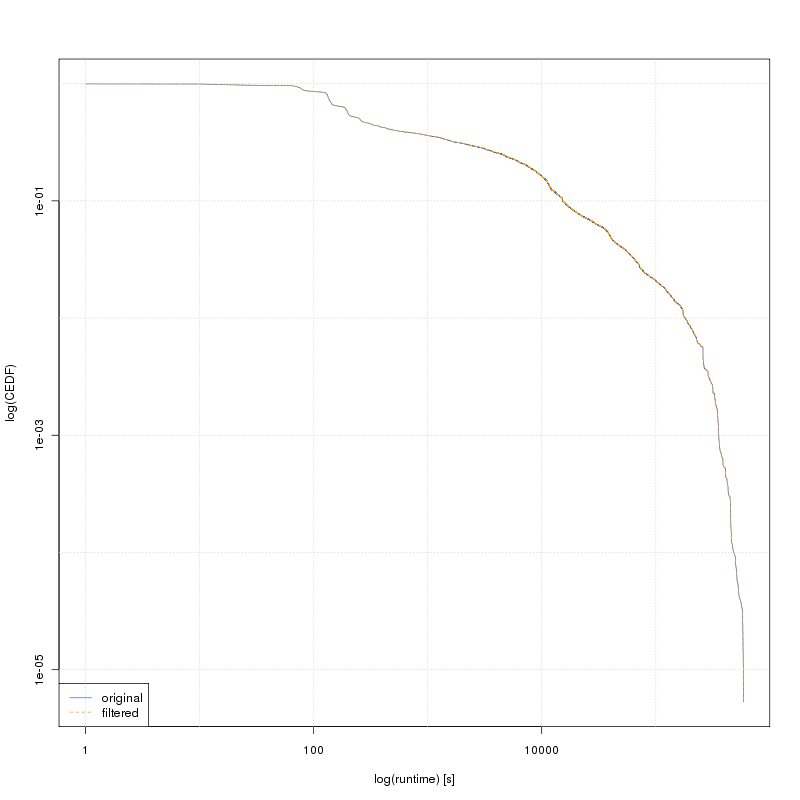}}
\quad
\subfloat[Q-Q plot]{\includegraphics[scale=0.17]{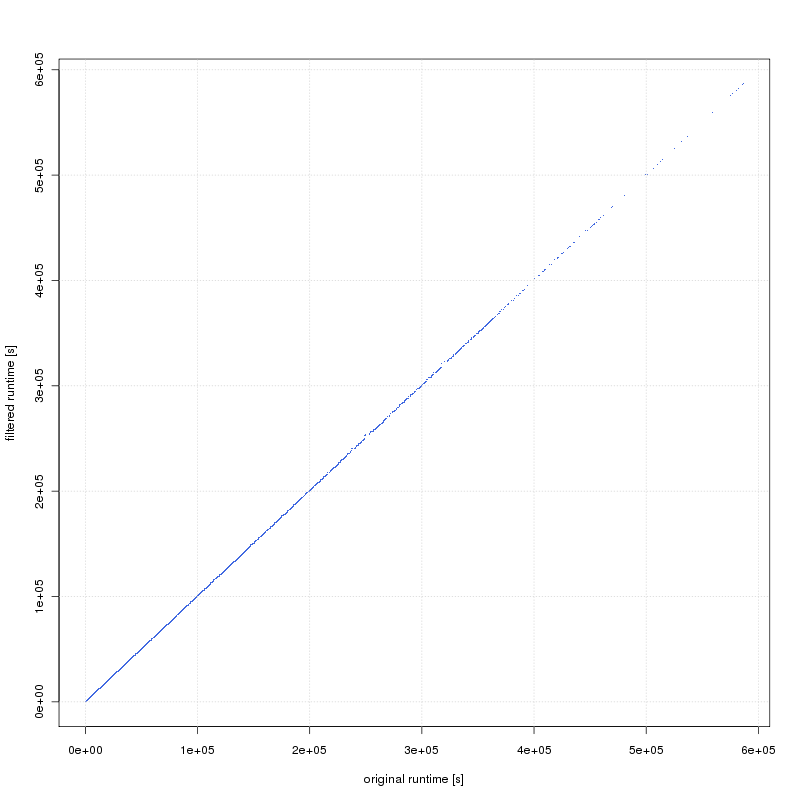}}
\caption{Comparison between original and filtered distributions of \texttt{runtime}.}
\label{fig:lcg-cmp-rt}
\end{figure*}

\begin{figure*}
\centering
\subfloat[Comparison of EDFs (in log-scale).]{\includegraphics[scale=0.17]{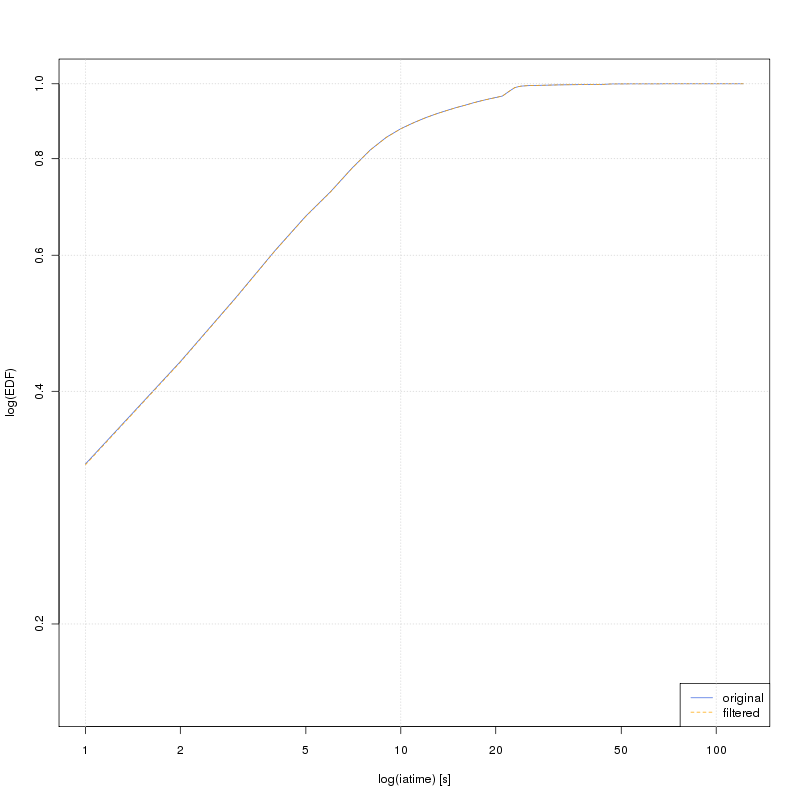}}
\quad
\subfloat[Comparison of CEDFs (in log-scale).]{\includegraphics[scale=0.17]{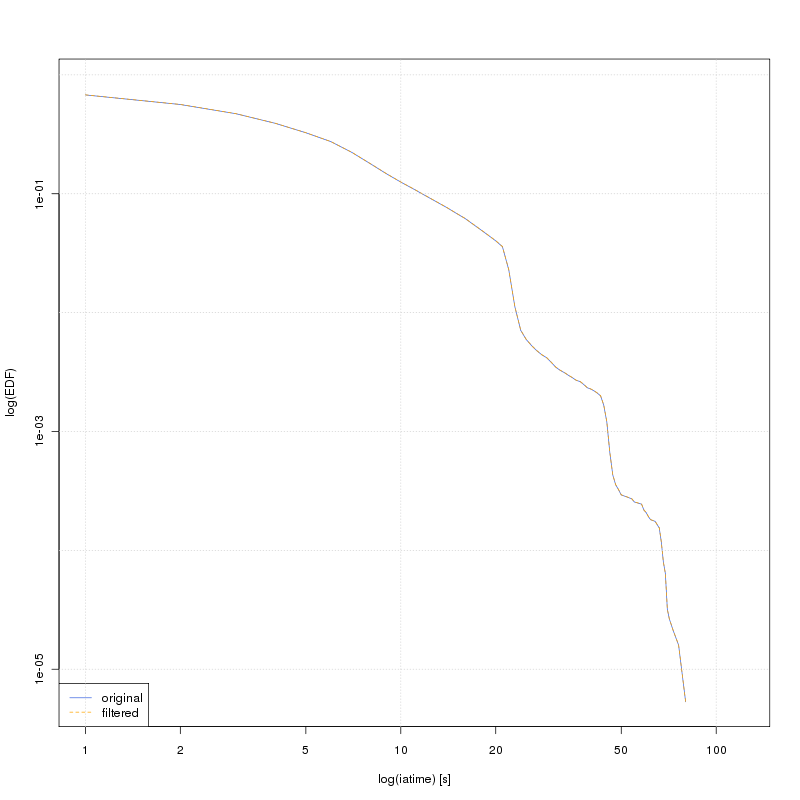}}
\quad
\subfloat[Q-Q plot.]{\includegraphics[scale=0.17]{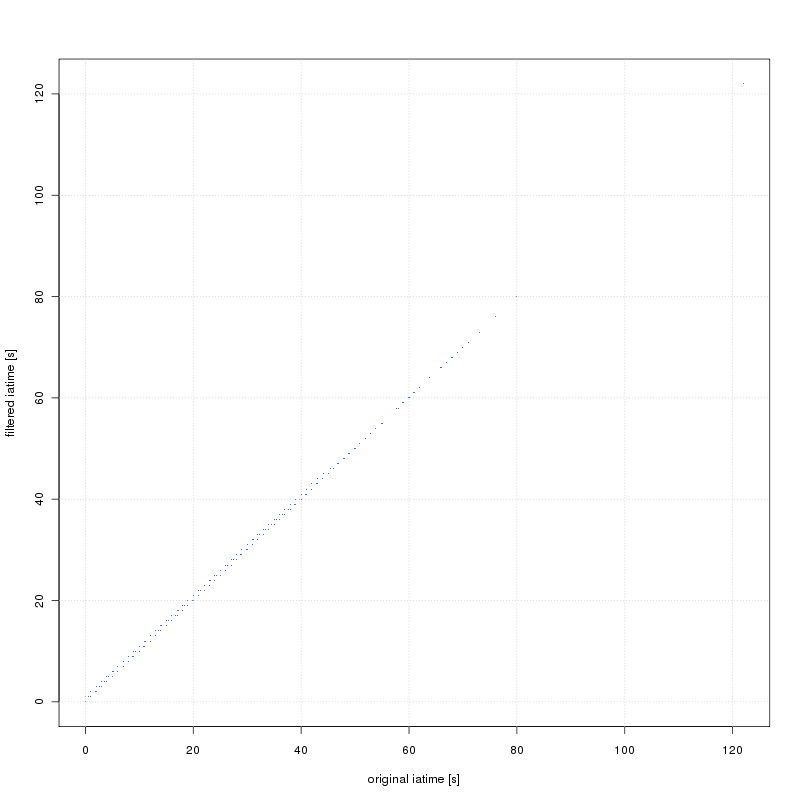}}
\caption{Graphical comparisons between original and filtered distributions of \texttt{iatime}.}
\label{fig:lcg-cmp-iat}
\end{figure*}

Looking at the data stored inside the trace, we find few entries with the attribute \texttt{runtime} set to $0$.
The number of these entries only represented the $0.28\%$ of the whole dataset.
We are not able to give them an exact interpretation; we can only suppose that their presence could represent an error condition (e.g., a failure during the job execution) or the run time of very fast jobs, whose execution times were smaller than the time granularity used for tracing the job execution.
Due to this uncertainty, we consider two versions of the trace, the original and the filtered version (i.e., without the entries with run times equal to zero), and performed for each of them a separated preliminary data analysis in order to understand the impact of the removal of these suspicious entries.
Since our aim is to characterize the interarrival time and the run time, we investigate for any difference in the distributions of both \texttt{runtime} and \texttt{iatime} attributes.
From \mgTabRef{lcg-cmp-stats}, it results that only the \texttt{runtime} distribution showed non negligible variations after the filtering process; indeed, the removal of suspicious observations, made to increase the mean value and to slightly move to the right the skewness of the distribution (i.e., to stretch its right tail).
However, subsequent statistical tests revealed that such modifications were not significant.
As a matter of fact, from a graphical comparison of the original and the filtered distribution of \texttt{runtime} (shown on \mgFigRef{lcg-cmp-rt}) and \texttt{iatime} (shown on \mgFigRef{lcg-cmp-iat}), we don't found any evident difference.
Moreover, further statistical hypothesis tests (shown on \mgTabRef{lcg-cmp-tests}) confirmed that the original and the filtered distribution of both attributes was not significantly different.~\footnote{We are unable to perform the Ansari-Bradley test due to integer overflow errors possibly caused by the large size of the data set.}
Since the distributions did not change significantly, we decide to remove from the trace all the entries with run time equal to $0$.

\begin{table*}
\centering
\caption{Summary statistics for the original and filtered distributions of \texttt{runtime} and \texttt{iatime}.}
\label{tab:lcg-cmp-stats}
\begin{tabular}{lrrrr}
\toprule
 & \multicolumn{2}{c}{\texttt{runtime}} & \multicolumn{2}{c}{\texttt{iatime}} \tabularnewline
 & \multicolumn{1}{c}{original} & \multicolumn{1}{c}{filtered} & \multicolumn{1}{c}{original} & \multicolumn{1}{c}{filtered} \tabularnewline
\midrule
\emph{Count} & $188041$ & $187520$ & $188041.00$ & $187520.00$ \tabularnewline
\emph{Min} & $0$ & $1$ & $0.00$ & $0.00$ \tabularnewline
\emph{1st Quartile} & $136$ & $136$ & $1.00$ & $1.00$ \tabularnewline
\emph{Median} & $255$ & $255$ & $3.00$ & $3.00$ \tabularnewline
\emph{Mean} & $8971$ & $8996$ & $5.05$ & $5.07$ \tabularnewline
\emph{3rd Quartile} & $4490$ & $4537$ & $7.00$ & $7.00$ \tabularnewline
\emph{Max} & $586700$ & $586700$ & $122.00$ & $122.00$ \tabularnewline
\bottomrule
\end{tabular}
\end{table*}


\begin{table*}
\centering
\caption{Statistical hypothesis tests, at $5\%$ of significance level, performed for assessing any significant difference between the original and filtered distributions of \texttt{runtime} and \texttt{iatime}: a non-rejected null hypothesis $\mathcal{H}_0$ indicates the original and filtered distributions are statistically equivalent.
The ``n/a'' symbol means ``not available'' and is due to integer overflow errors during the computation.}
\label{tab:lcg-cmp-tests}
\begin{tabular}{lrcrc}
\toprule
 & \multicolumn{2}{c}{\texttt{runtime}} & \multicolumn{2}{c}{\texttt{iatime}} \tabularnewline
 & \multicolumn{1}{c}{$p$-value} & $\mathcal{H}_0$ & \multicolumn{1}{c}{$p$-value} & $\mathcal{H}_0$ \tabularnewline
\midrule
\emph{Kolmogorov-Smirnov test} & $0.4669$ & no reject & $0.5750$ & no reject \tabularnewline
\emph{Pooled (Welch) $t$-test} & $0.8162$ & no reject & $0.2006$ & no reject \tabularnewline
\emph{Mann-Whitney test} & $0.1414$ & no reject & $0.1753$ & no reject \tabularnewline
\emph{$F$-test} & $0.5781$ & no reject & $0.4387$ & no reject \tabularnewline
\emph{Ansari-Bradley test} & n/a & n/a & n/a & n/a \tabularnewline
\midrule
\emph{Result} & \multicolumn{2}{c}{original $\equiv$ filtered} & \multicolumn{2}{c}{original $\equiv$ filtered} \tabularnewline
\bottomrule
\end{tabular}
\end{table*}

\subsection{Statistical Analysis of Workload Characteristics} \label{ssec:exp-stats}

\begin{figure*}
\centering
\subfloat[Mass-Count disparity for \texttt{iatime} attribute.]{\includegraphics[scale=0.25]{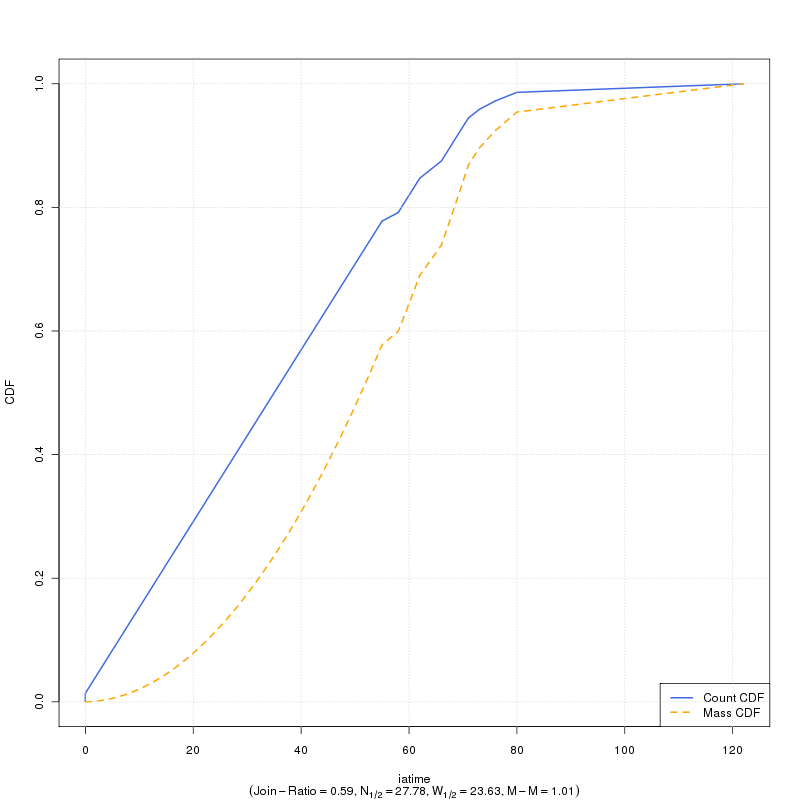}}
\quad
\subfloat[Mass-Count disparity for \texttt{runtime} attribute.]{\includegraphics[scale=0.25]{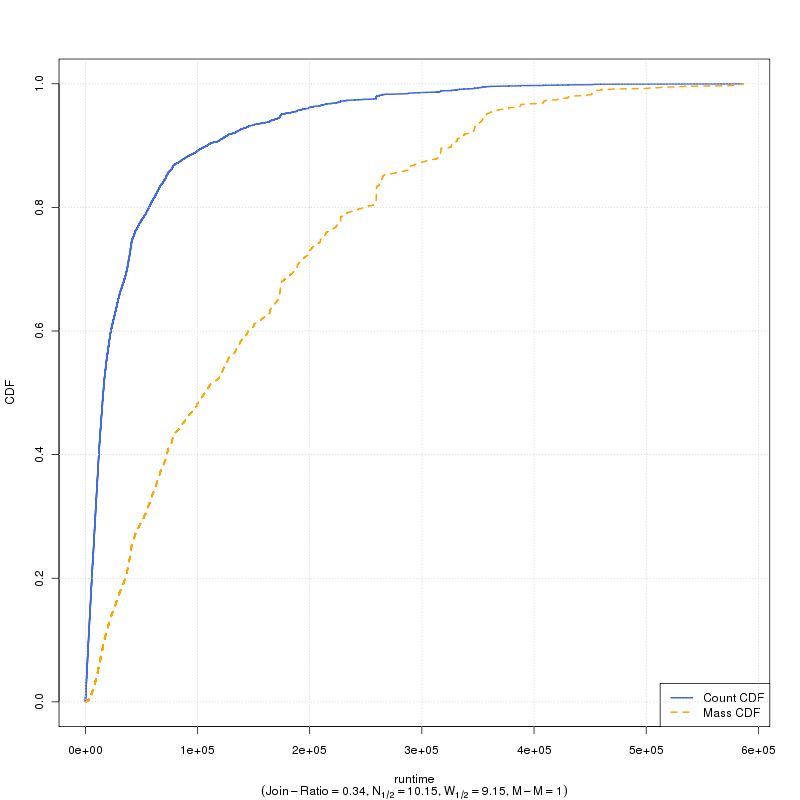}}
\caption{Mass-Count disparity plots for \texttt{iatime} and \texttt{runtime} attributes: the \texttt{runtime} distribution present an indication of the power-law property.}
\label{fig:lcg-eda_powerlaw_ineq_mc}
\end{figure*}

In \mgTabRef{lcg-sumstats} are shown summary statistics for the filtered distribution of \texttt{iatime}, \texttt{runtime} \texttt{vo} and \texttt{uid} attributes at grid level.
It is important to note that, since the \texttt{vo} and \texttt{uid} attributes are categorical, we consider only, as summary statistics, the order statistics (i.e., quantile values) and the mode of the related frequency distribution; in the table, the symbol ``n/a'', which stands for ``not applicable'', is used to indicate that the computation of the associated statistics is meaningless.
Analyzing the values of skewness and kurtosis, shown in the above table, we can conclude that both the distribution of \texttt{iatime} and that of \texttt{runtime} is highly positive (i.e., right) skewed and presents a high and thin peak around the mean; this is a possible indication of a power-law distribution.
This is even confirmed by the high (i.e., greater than $1$) sample coefficient of variation (CV), that is the ratio between the sample standard deviation and the sample mean, and by the sample mean (i.e., the center of the mass), which is far from the sample median (i.e., the center of the distribution); indeed, the mass-count disparity plot, shown in \mgFigRef{lcg-eda_powerlaw_ineq_mc}, reveals that the distribution of \texttt{runtime} attribute is likely to be a power-law distribution; the skewness is particularly strong for the distribution of \texttt{runtime}.
\begin{table*}
\centering
\caption{Summary statistics for the \texttt{iatime}, \texttt{runtime}, \texttt{vo} and \texttt{uid} attributes (``n/a'' stands for ``not applicable'').}
\label{tab:lcg-sumstats}
\begin{small}
\begin{tabular}{lrrcc}
\toprule
& \texttt{iatime} & \texttt{runtime} & \texttt{vo} & \texttt{uid} \\
\midrule
Count & $187520.000$ & $187520.000$ & $28$ & $216$ \\
Min & $0.000$ & $1.000$ & dzero, photon & \multicolumn{1}{p{3cm}}{102, 111, 112, 121, 133, 166, 182, 190, 206, 208, 217, 218, 229, 234, 252, 276, 294, 365, 373, 401, 426, 446, 447, 451, 454, 459} \\
$1^{st}$ Quartile & $1.000$ & $136.000$ & vlefi, vlibu & 22, 28, 80, 85, 441 \\
Median & $3.000$ & $255.000$ & babar, see & 51, 65, 74 \\
$3^{rd}$ Quartile & $7.000$ & $4537.000$ & phicos, zeus & 64, 123 \\
Maximum & $122.000$ & $586702.000$ & lhcb & 26 \\
Mode & $1.000$ & $134.000$ & lhcb & 26 \\
Mean & $5.068$ & $8995.843$ & n/a & n/a \\
Mean $95\%$ C.I. & $[5.041,5.095]$ & $[8847.045,9144.640]$ & n/a & n/a \\
Std Dev & $5.939$ & $32875.240$ & n/a \\
Std Dev $95\%$ C.I. & $[5.920,5.958]$ & $[32770.370,32980.800]$ & n/a & n/a \\
CV & $1.172$ & $3.654$ & n/a & n/a \\
IQR & $6.000$ & $4401.000$ & n/a & n/a \\
Kurtosis & $13.160$ & $65.765$ & n/a & n/a \\
Skewness & $2.432$ & $7.145$ & n/a & n/a \\
Quartile Skewness & $0.333$ & $0.946$ & n/a & n/a \\
\bottomrule
\end{tabular}
\end{small}
\end{table*}

\begin{figure*}
\centering
\subfloat[Run-sequence plot of \texttt{iatime} attribute.]{\includegraphics[scale=0.13]{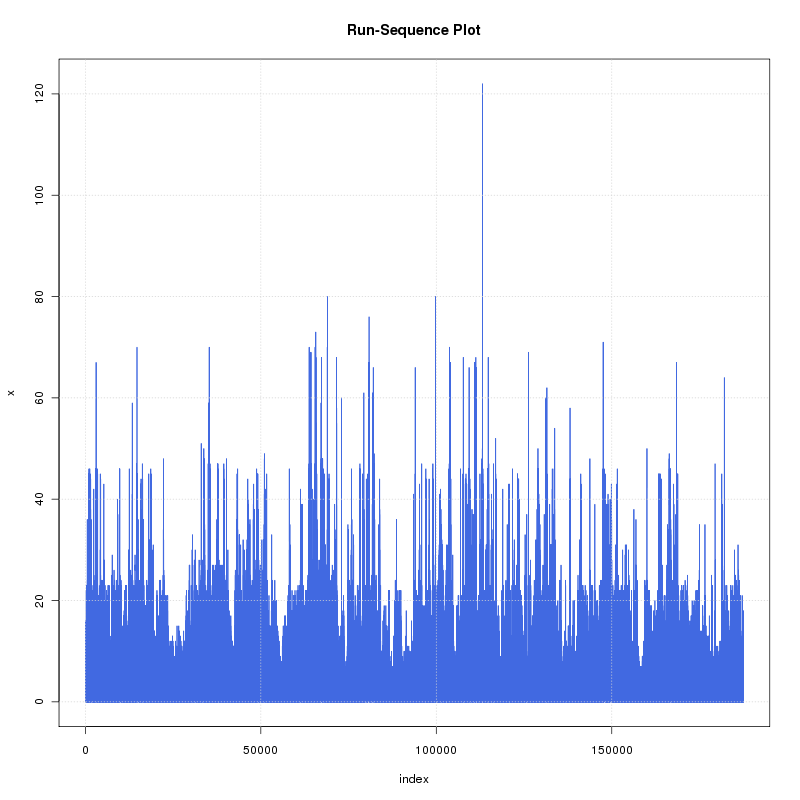}\label{fig:lcg-acor-iat_rt-iat-rs}}
\quad
\subfloat[ACF plot of \texttt{iatime} attribute.]{\includegraphics[scale=0.13]{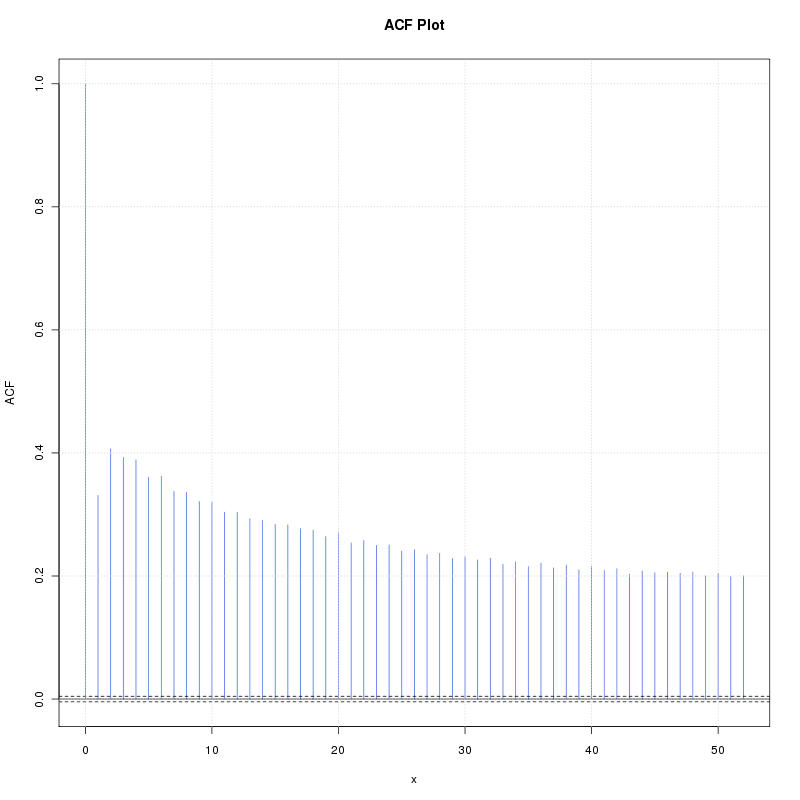}\label{fig:lcg-acor-iat_rt-iat-acf}}
\quad
\subfloat[Run-sequence plot of \texttt{runtime} attribute.]{\includegraphics[scale=0.13]{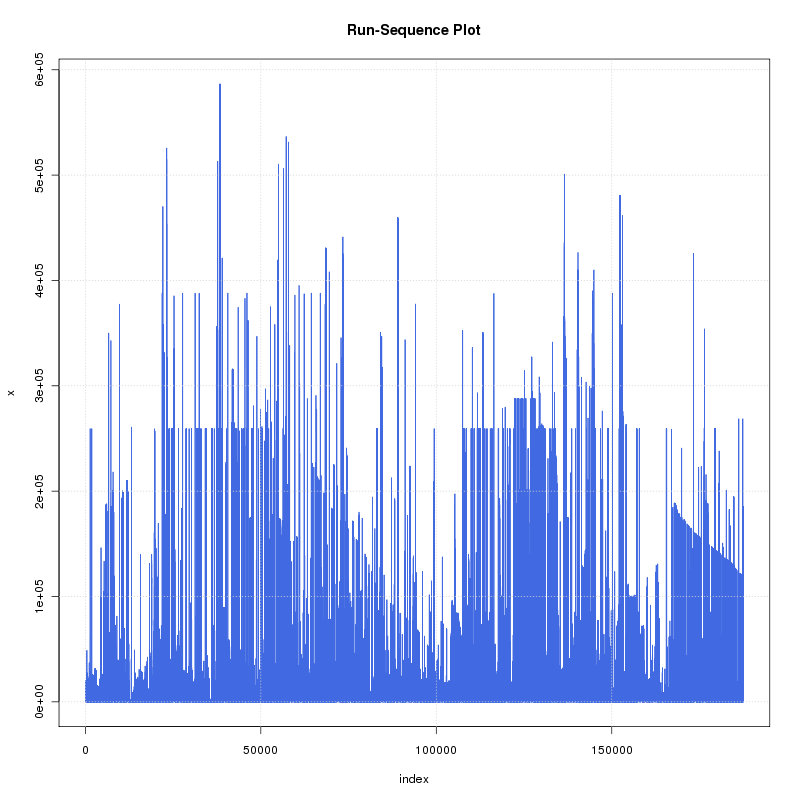}\label{fig:lcg-acor-iat_rt-rt-rs}}
\quad
\subfloat[ACF plot of \texttt{runtime} attribute.]{\includegraphics[scale=0.13]{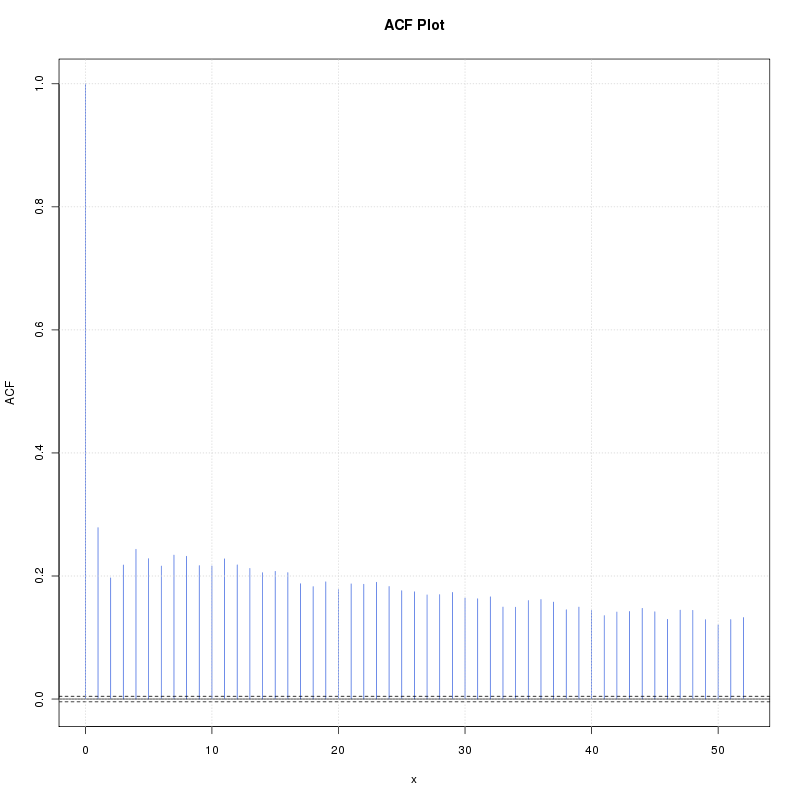}\label{fig:lcg-acor-iat_rt-rt-acf}}
\caption{Run-sequence plot and autocorrelation function (ACF) plot for the \texttt{iatime} and \texttt{runtime} attributes.}
\label{fig:lcg-acor-iat_rt}
\end{figure*}

\begin{figure*}
\centering
\includegraphics[scale=0.45]{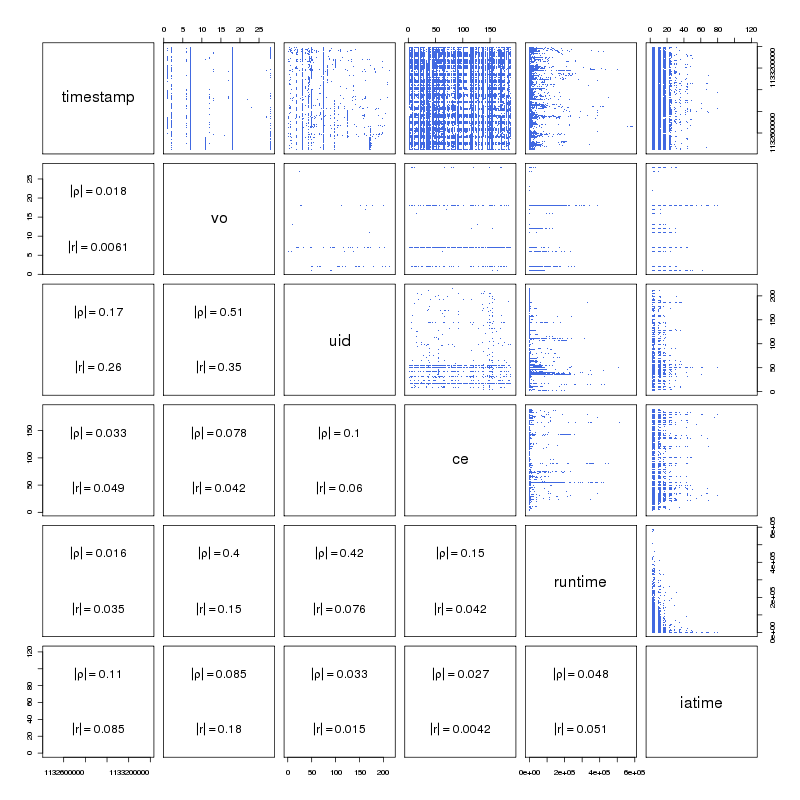}
\caption{Scatter-plot matrix among the workload attributes: $|\rho|$ and $|r|$ are the absolute values of the Spearman's and Pearson's correlation coefficients, respectively.}
\label{fig:lcg-scattermat}
\end{figure*}

\begin{figure*}
\centering
\subfloat[Number of jobs versus job runtime, at VO level.]{\includegraphics[scale=0.25]{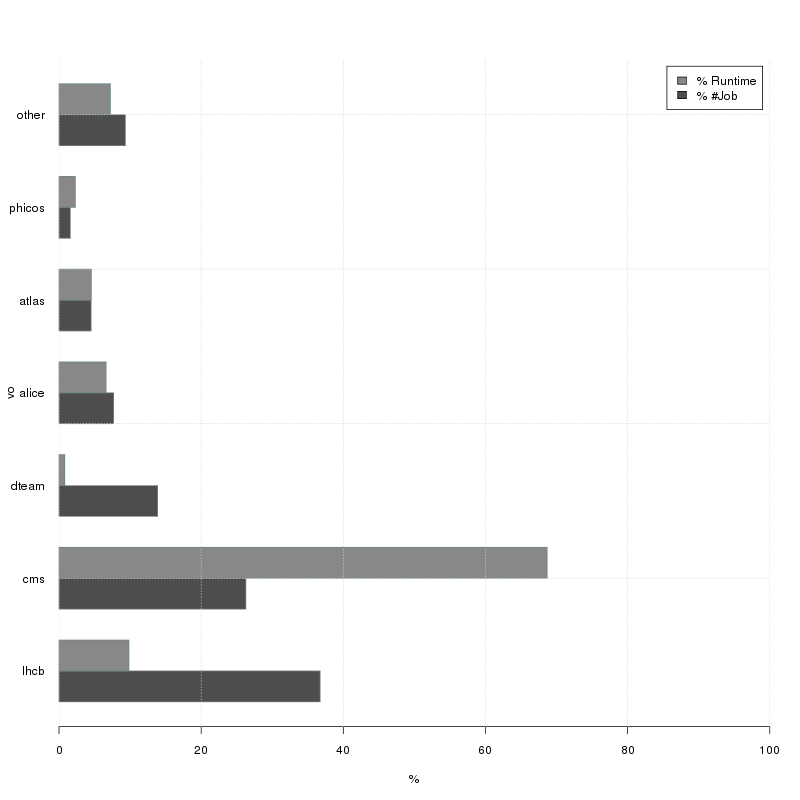}\label{fig:lcg-vo-njobs_rt}}
\quad
\subfloat[Number of jobs versus job runtime, at user level.]{\includegraphics[scale=0.25]{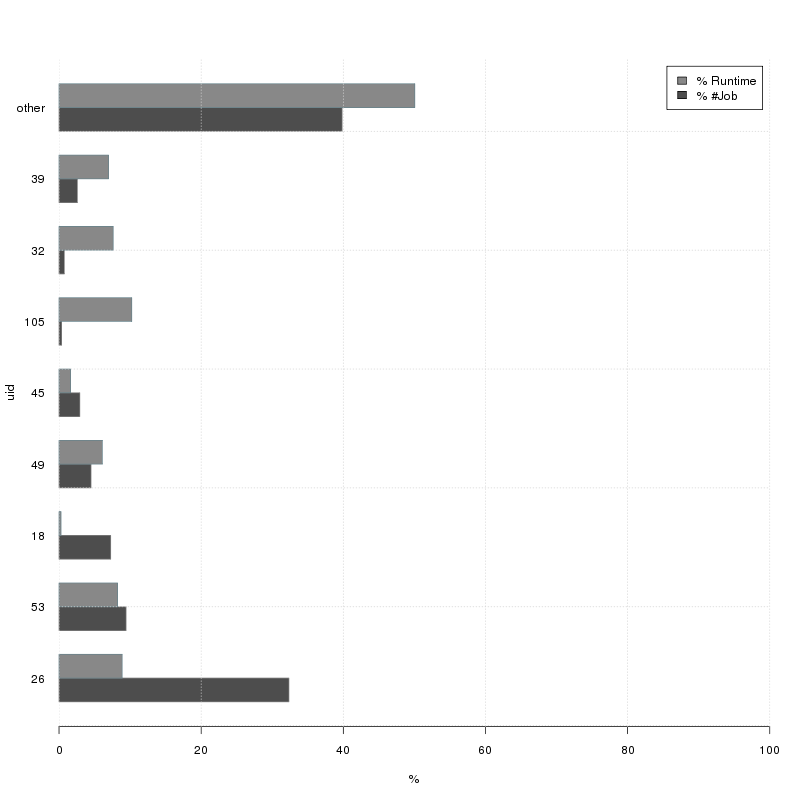}\label{fig:lcg-uid-njobs_rt}}
\caption{Number of jobs versus job runtime, at both VO and user level.}
\label{fig:lcg-njobs_rt}
\end{figure*}

In order to discover correlations among workload characteristics, we use both graphical and statistical tools.
In \mgFigRef{lcg-acor-iat_rt} are presented the run-sequence plot and the autocorrelation plot for the \texttt{iatime} and \texttt{runtime} attributes.
From the run-sequence plot of the \texttt{iatime} attribute (see \mgFigRef{lcg-acor-iat_rt-iat-rs}) we can note that the trend is very bursty, while from the ACF plot (see \mgFigRef{lcg-acor-iat_rt-iat-acf}) we can observe the possible presence of a weak short-range dependence and of a moderate long-range dependence.
For what regards the \texttt{runtime} attribute, its run-sequence plot (see \mgFigRef{lcg-acor-iat_rt-rt-rs}) shows a bursty behavior too; however, the ACF plot (see \mgFigRef{lcg-acor-iat_rt-rt-acf}) shows that the presence of both short-range dependence and long-range dependence is very weak.
For what concern the correlation among different attributes, we use the scatter-matrix plot.
As shown in \mgFigRef{lcg-scattermat}, at grid level we don't find any significant linear correlation among the attributes, since the absolute values of the Pearson's correlation coefficient $r$ are all nearly equal to $0$; among these values, we can note that there might be a weak linear correlation between the \texttt{timestamp} and \texttt{uid} attributes ($|r|=0.28$), the \texttt{vo} and \texttt{uid} attributes ($|r|=0.35$), and the \texttt{vo} and \texttt{runtime} attributes ($|r|=0.15$).
Note that, while there is some kind of correlation between \texttt{uid} and \texttt{timestamp}, it disappears if we concentrate on interarrival times (attribute \texttt{iatime}).
Instead, looking at the absolute values of the Spearman's correlation coefficient $\rho$, a much stronger and possibly non-linear correlation seems to be present between the \texttt{vo} and \texttt{uid} attributes ($|\rho|=0.51$), the \texttt{vo} and \texttt{runtime} attributes ($|\rho|=0.4$), and the \texttt{uid} and \texttt{runtime} attributes ($|\rho|=0.42$).
This figure reveals us other important facts.
Firstly, from the \texttt{timestamp}-\texttt{vo} and \texttt{timestamp}-\texttt{uid} plots, we can note that there are VOs and users, respectively, which submit more jobs than others.
Secondly, from the \texttt{runtime}-\texttt{iatime} plot, it seems there is little difference between the interarrival time of long jobs and the one of short jobs, and the majority of jobs are submitted at short or moderate interarrival time.
These considerations suggest that additional investigations on these relationships are to be carried on.
We try to understand what are the most influential VOs and users with respect to the number of jobs submitted and the runtime; the result is reported in \mgFigRef{lcg-njobs_rt}.
To obtain this information, we create the distribution of the number of submitted jobs and the one of the job runtimes, for both VOs and users, and for every distribution (of each attribute) we take the $5$ most influential attributes.
Specifically, we consider the five VOs with the greatest number of submitted jobs and the (possibly different) five VOs with the greatest execution times, and we join them together (removing the overlapping ones).
The result is shown in \mgFigRef{lcg-vo-njobs_rt}; from the figure we can note that the five most influential VOs in terms of number of submitted jobs are (in order of importance): \emph{lhcb}, \emph{cms}, \emph{dteam}, \emph{alice} and \emph{atlas}; while the five ones with the longest runtime are (in order of importance): \emph{cms}, \emph{lhcb}, \emph{alice}, \emph{atlas} and \emph{phicos}; the label \emph{other} is used to aggregate the values of all the other least influential VOs.
Similarly, from \mgFigRef{lcg-uid-njobs_rt} we can observe the most influential users in terms of number of submitted jobs; they are (in order of importance): $26$, $53$, $18$, $49$ and $45$; while the five ones with the greatest runtime are (in order of importance): $105$, $26$, $53$, $32$ and $39$; as for the attribute \texttt{vo}, the label \emph{other} is used to aggregate the values of all the other least influential users.
Further investigations revealed that: (1) nearly each user belongs to only $1$ VO but $12$ users, who belong from $2$ to $4$ different VOs; in particular, $3$ of such users (i.e., user $45$, $53$, and $105$) are among the most influential users and nearly all of them belong to the most influential VOs , and (2) roughly all the most influential VOs (i.e., \emph{atlas}, \emph{cms}, \emph{dteam}, \emph{lhcb}) group the largest number of users.
Consequently, we can conclude that the most influential VOs, namely \emph{alice}, \emph{atlas}, \emph{cms}, \emph{dteam} and \emph{lhcb}, are, at the same time, those with the largest user community, those with the largest number of submitted jobs, those with the longest job runtime; further, nearly all of the influential users belongs to such VOs.

\subsection{Cluster-Based Model for Workload Characteristics} \label{ssec:exp-model}

In this section we present an approach to model workload characteristic primarily based on cluster analysis.
Among the attributes in the LCG trace, the ones that play a central role from the point of view of a grid scheduler are the job interarrival time (i.e., \texttt{iatime} attribute), the user who submitted the job (i.e., \texttt{uid} attribute) and the job execution time (i.e., \texttt{runtime} attribute).
The knowledge of the interarrival time lets the scheduler to predict when the next job will likely arrive; the information on the job execution time lets the scheduler to predict the possible amount of time a job will execute and thus to assign to it a suitable computing machine.
Finally, the knowledge about the user allows to characterize typical user usage patterns.
It results that, in order to create a realistic workload model, it is important to characterize all these three workload characteristics.

As discussed in \mgSSecRef{method-model} our intent was to create a workload model $\mathcal{M}=\langle\mathcal{C},\mathbf{u},\mathcal{W},\mathbf{S}\rangle$ where $\mathcal{C}=\{C_1,\ldots,C_m\}$ is the set of cluster found through cluster analysis (summarizing pairs of interarrival and run times), $\mathbf{u}$ is the user weights vector (giving the fraction of job submissions for each user), $\mathcal{W}=\{\mathbf{w}_1,\ldots,\mathbf{w}_n\}$ is the set of vectors of user workload contributions (representing the weights that each user provides inside each cluster), and $\mathbf{S}=(s_{ij})$ is the $m \times 2$ matrix containing information on the variability of every attribute $j$ inside each cluster $C_i$.

\begin{figure}
\centering
\subfloat[$1^{\text{st}}$ sample.]{\includegraphics[scale=0.11]{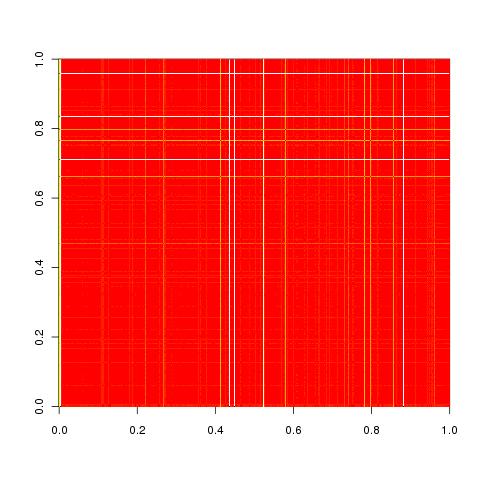}}
\quad
\subfloat[$2^{\text{nd}}$ sample.]{\includegraphics[scale=0.11]{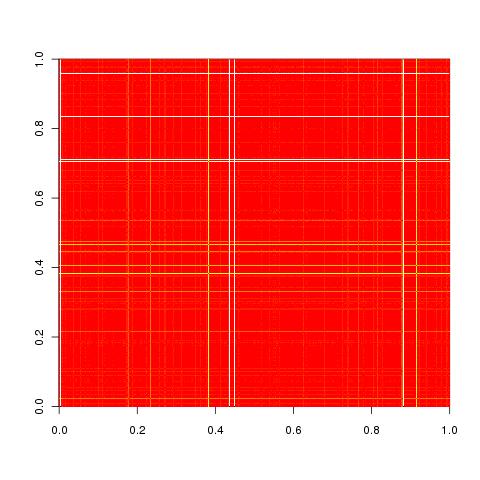}}
\quad
\subfloat[$3^{\text{rd}}$ sample.]{\includegraphics[scale=0.11]{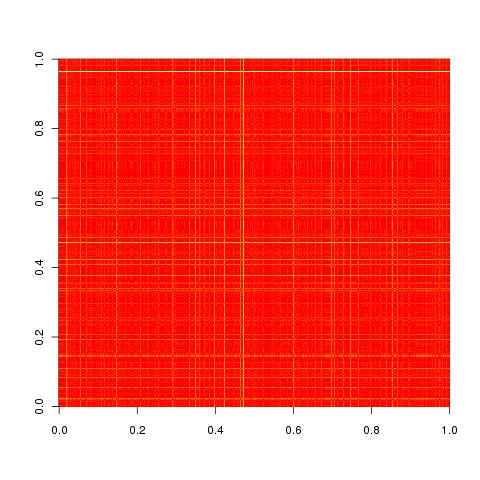}}
\quad
\subfloat[$4^{\text{th}}$ sample.]{\includegraphics[scale=0.11]{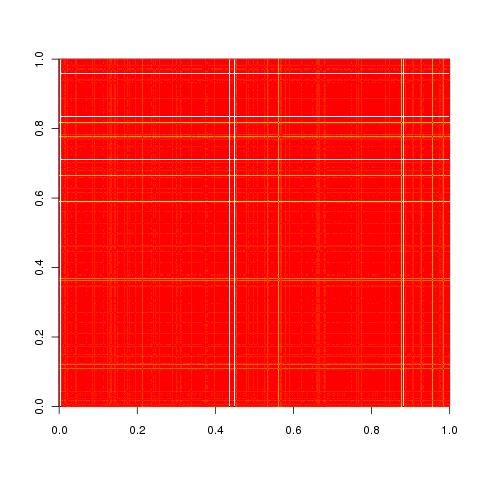}}
\quad
\subfloat[$5^{\text{th}}$ sample.]{\includegraphics[scale=0.11]{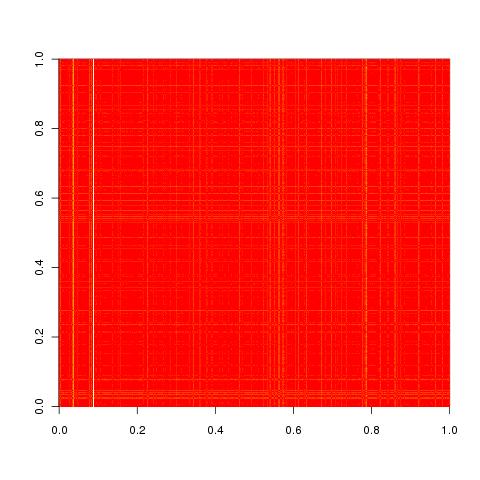}}
\quad
\subfloat[$6^{\text{th}}$ sample.]{\includegraphics[scale=0.11]{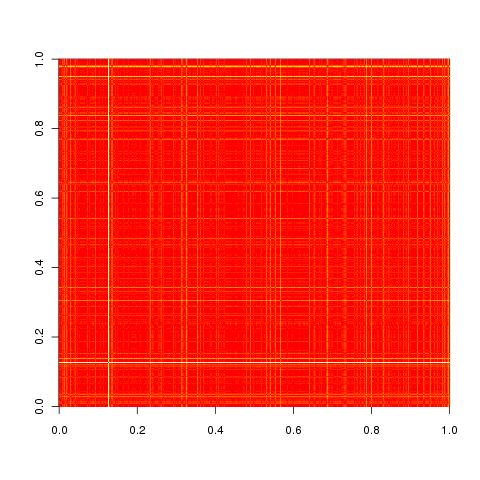}}
\quad
\subfloat[$7^{\text{th}}$ sample.]{\includegraphics[scale=0.11]{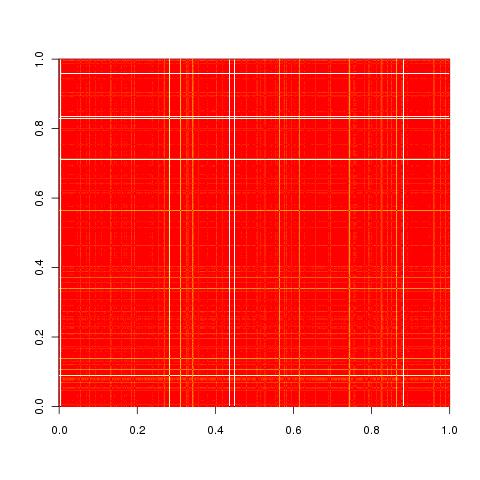}}
\\
\subfloat[$8^{\text{th}}$ sample.]{\includegraphics[scale=0.11]{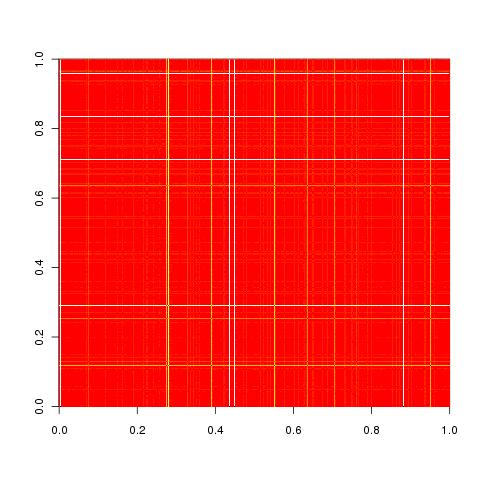}}
\quad
\subfloat[$9^{\text{th}}$ sample.]{\includegraphics[scale=0.11]{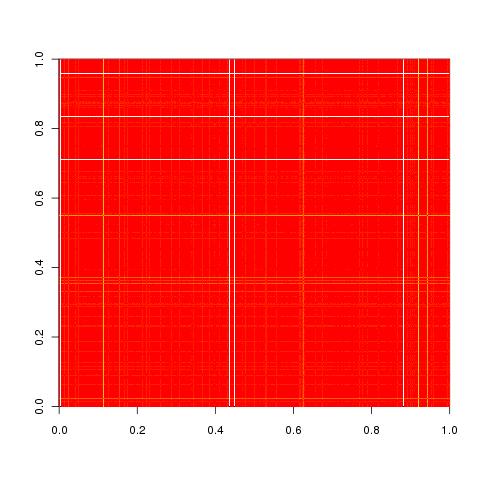}}
\quad
\subfloat[$10^{\text{th}}$ sample.]{\includegraphics[scale=0.11]{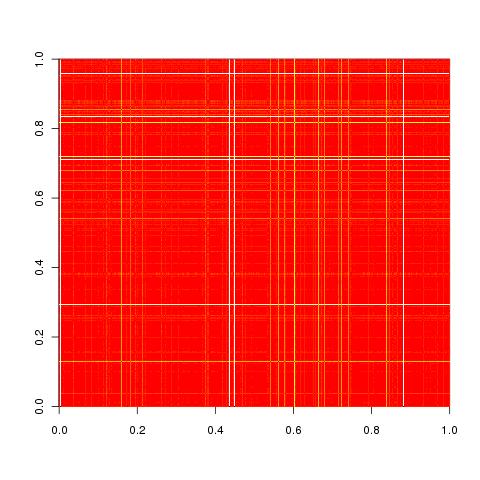}}
\quad
\subfloat[$11^{\text{th}}$ sample.]{\includegraphics[scale=0.11]{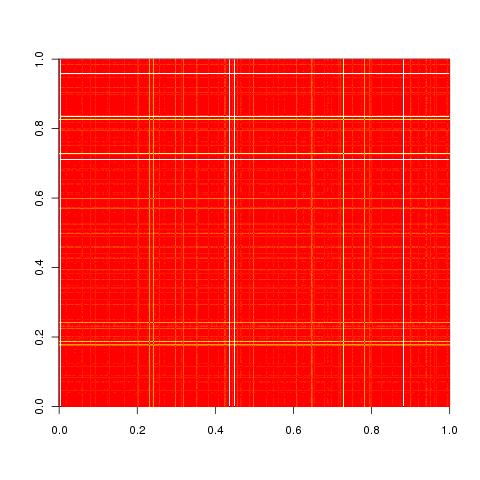}}
\quad
\subfloat[$12^{\text{th}}$ sample.]{\includegraphics[scale=0.11]{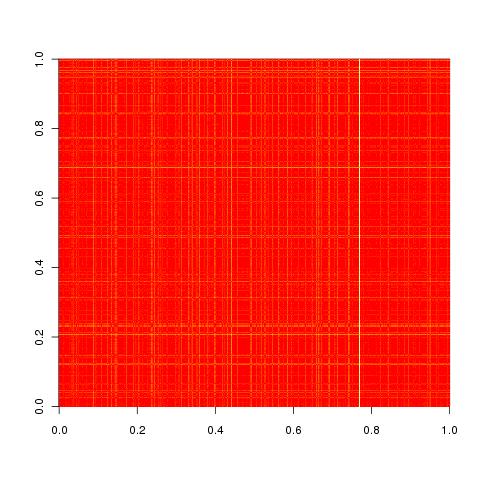}}
\quad
\subfloat[$13^{\text{th}}$ sample.]{\includegraphics[scale=0.11]{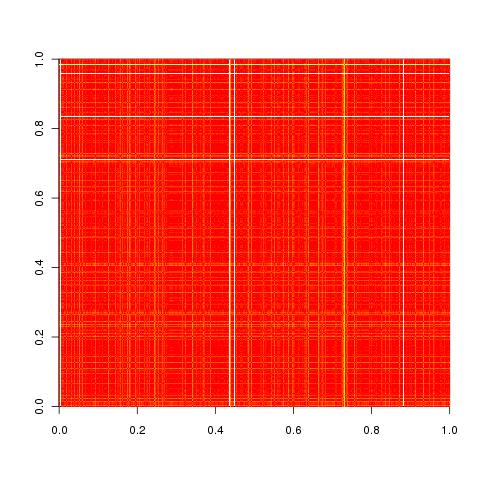}}
\quad
\subfloat[$14^{\text{th}}$ sample.]{\includegraphics[scale=0.11]{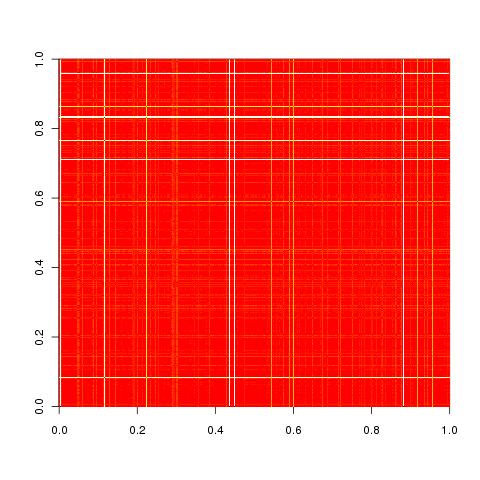}}
\\
\subfloat[$15^{\text{th}}$ sample.]{\includegraphics[scale=0.11]{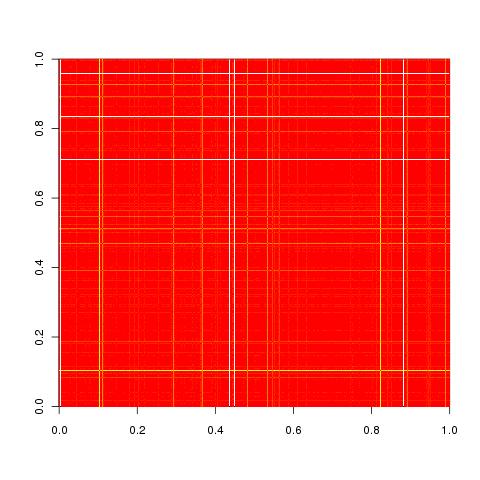}}
\quad
\subfloat[$16^{\text{th}}$ sample.]{\includegraphics[scale=0.11]{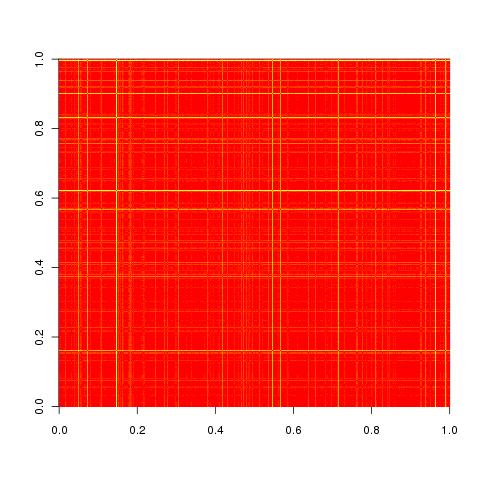}}
\quad
\subfloat[$17^{\text{th}}$ sample.]{\includegraphics[scale=0.11]{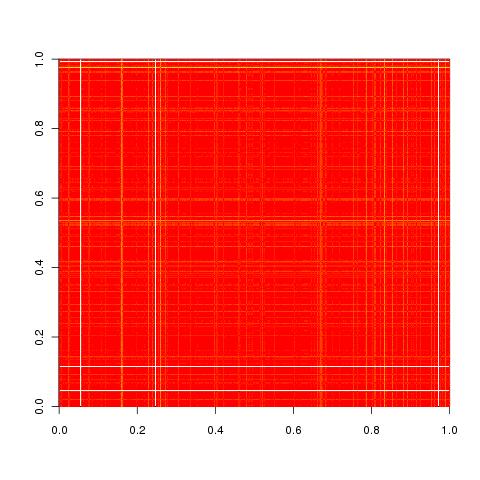}}
\quad
\subfloat[$18^{\text{th}}$ sample.]{\includegraphics[scale=0.11]{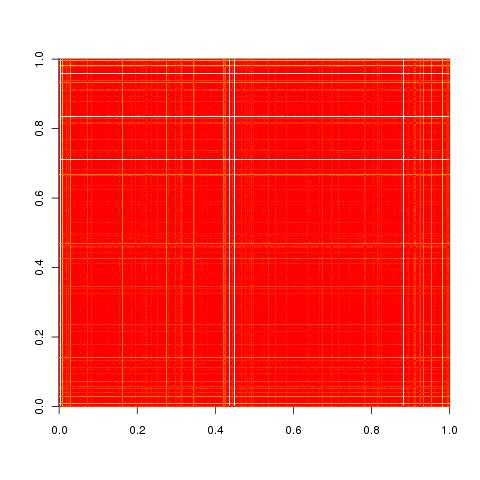}}
\quad
\subfloat[$19^{\text{th}}$ sample.]{\includegraphics[scale=0.11]{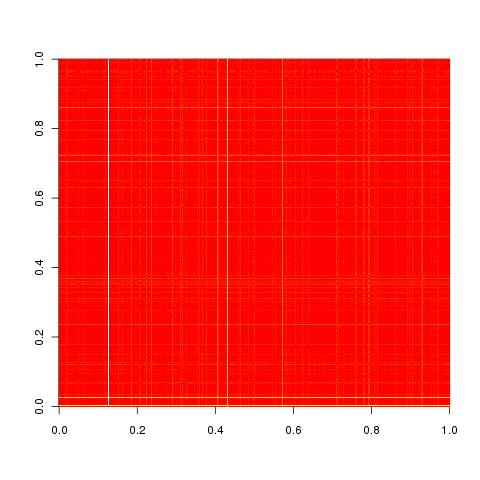}}
\quad
\subfloat[$20^{\text{th}}$ sample.]{\includegraphics[scale=0.11]{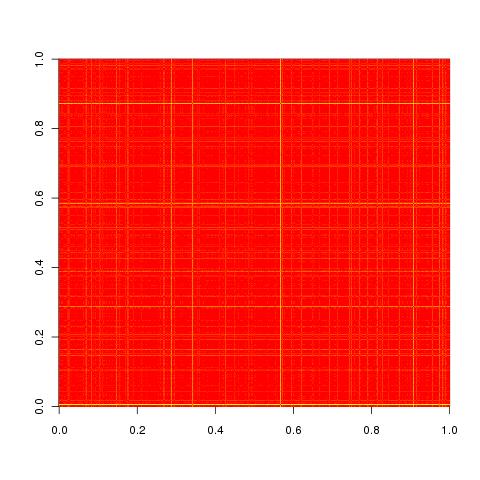}}
\quad
\subfloat[$21^{\text{st}}$ sample.]{\includegraphics[scale=0.11]{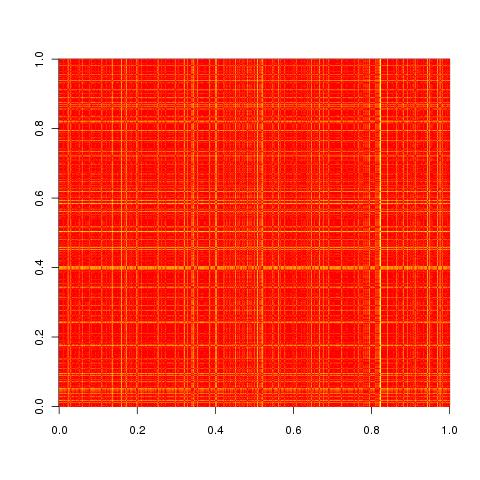}}
\caption{VAT images for the first $21$ data samples of size $1875$. From each VAT image it seems there is a possible lacks of clustering tendency.}
\label{fig:lcg-iat_rt-cluster-vat}
\end{figure}

Before applying any cluster algorithm, we firstly normalize our data set and then investigated on the clustering tendency.
To this purpose, we use the VAT graphical method and the Hopkins test.
Unfortunately, both methods incur in the curse of dimensionality problem and hence we are not able to apply them on the entire data set.
Nevertheless, we try to apply these methods on a restricted data set, anyway; specifically, we reduce the size of the original data set by taking only $1875$ observations (approximatively $1\%$ of the whole data set) for VAT and $938$ observations (approximatively $0.5\%$ of the entire data set) for the Hopkins test.
It is important to note that this reduction of size might remove patterns that are important for detecting the possible natural structure of data; in order to try to capture these patterns as much as possible, we apply the clustering tendency tests on $30$ different random data sample.
In \mgFigRef{lcg-iat_rt-cluster-vat} is presented the result of VAT for the first $21$ samples (the other samples exhibited similar behavior).
From the figure, we can conclude that apparently there is no indication of clustering tendency; however, this result should be taken with care since it is based only on a small fraction of the entire data.
For what concern the Hopkins test, we use, for each data sample, a sampling windows of size $94$ (approximatively $10\%$ of the data sample size); for each data sample, we run the Hopkins test against $5$ sampling windows and taken the average value as the final value of the Hopkins statistics. 
The result was that for every data sample we obtain an average Hopkins statistic nearly equal to $1$.
Thus, according to the Hopkins test, there would be an high confidence on the presence of natural clusters in the data set.
This result is totally in contrast with the one we obtain with VAT.
Again, due to the reduction in size of data set, this finding should be taken with care.

Since we are not able to prove neither the presence nor the absence of natural clustering tendency, we apply some cluster algorithm, anyway.
As reported in \mgSSecRef{method-model}, we consider three different partitional algorithms, namely CLARA, MCLUST and DBSCAN.
The CLARA algorithm requires only one mandatory parameter, namely the number of cluster to look for.
In our experiments, we run CLARA on the whole data set using the Euclidean distance and by trying different number of clusters, from a minimum of $2$ to a maximum of $11$; for the remaining input parameters, namely the sample size and the number of samplings, we employ the same values used in the original algorithm, that is, $40+2k$ for the sample size (where $k$ is the current number of clusters), and $5$ for the number of samplings.
In order to evaluate the best number of clusters for CLARA, we use the silhouette coefficient.
As can be noted from \mgTabRef{exp-cluster-clara}, the highest silhouette coefficient is $0.8984573$ and it is obtained with $2$ clusters.
According to the interpretation of the silhouette coefficient provided in \mgSSecRef{method-model}, the found clustering structure is strong.
\begin{table}
\centering
\caption{Clustering with CLARA: average silhouette coefficient against different number of clusters.}
\label{tab:exp-cluster-clara}
\begin{tabular}{rr}
\toprule
\multicolumn{1}{c}{Number of Clusters} & \multicolumn{1}{c}{Silhouette Coefficient} \\
\midrule
2 & 0.8984573 \\
3 & 0.6093906 \\
4 & 0.5970364 \\
5 & 0.4900663 \\
6 & 0.5204294 \\
7 & 0.4925825 \\
8 & 0.4776292 \\
9 & 0.4929831 \\
10 & 0.5360077 \\
11 & 0.5368265 \\
\bottomrule
\end{tabular}
\end{table}

The MCLUST algorithm requires only one mandatory parameter, namely the maximum number of clusters to look for (with a minimum of $2$).
In our experiments, we instruct MCLUST to use the value $10$ as the maximum number of clusters.
When we first run this algorithm, we face with the curse of dimensionality problem; indeed, MCLUST does not scale well for large data set.
Similarly to what we have done for clustering tendency, to mitigate this problem we reduce the size of the data set to $18752$ (i.e., to $10\%$ of the total number of observations); moreover, in order to capture as much as possible important data patterns, we apply MCLUST on $10$ different random data samples.
The results of MCLUST for every run are reported on \mgTabRef{exp-cluster-mclust}.
These results are misleading; indeed, with different data sample we obtain very different values of BIC; furthermore, the fact that for every run we obtain the maximum number of clusters that we set as input to MCLUST, should be judged with suspect.
For these reasons we decide to do not consider the results obtained with MCLUST.
\begin{table}
\centering
\caption{Clustering with MCLUST: BIC and number of clusters obtained with different data samples of size $18752$.}
\label{tab:exp-cluster-mclust}
\begin{tabular}{rrr}
\toprule
\multicolumn{1}{c}{Run} & \multicolumn{1}{c}{BIC} & \multicolumn{1}{c}{Number of Clusters} \\
\midrule
1 & 41387.73 & 10 \\
2 & 41152.10 & 10 \\
3 & 31619.22 & 10 \\
4 & 40048.53 & 10 \\
5 & 14519.82 & 10 \\
6 & 15455.38 & 10 \\
7 & -787.22 & 10 \\
8 & 43141.89 & 10 \\
9 & 44458.10 & 10 \\
10 & 22925.48 & 10 \\
\bottomrule
\end{tabular}
\end{table}

The DBSCAN algorithm requires two mandatory parameters, namely the radius $\epsilon$ (used for limiting the neighborhood of each point) and the threshold \emph{MinPts} (representing the minimum number of points inside the neighborhood of a centroid).
In our experiments, we set $\epsilon$ to $0.1053428$ and \emph{MinPts} to $5$.
The result of DBSCAN was a total of $242$ clusters.
As mentioned in \mgSSecRef{method-model} we do not find any clustering validation measure suitable for density-based cluster algorithms; thus, being unable to quantitatively evaluate the goodness of this result, we decide to not consider it.

\begin{figure}
\centering
\subfloat[The CLUSPLOT: the first $2$ principal components explain $100\%$ of the point variability.]{\includegraphics[scale=0.25]{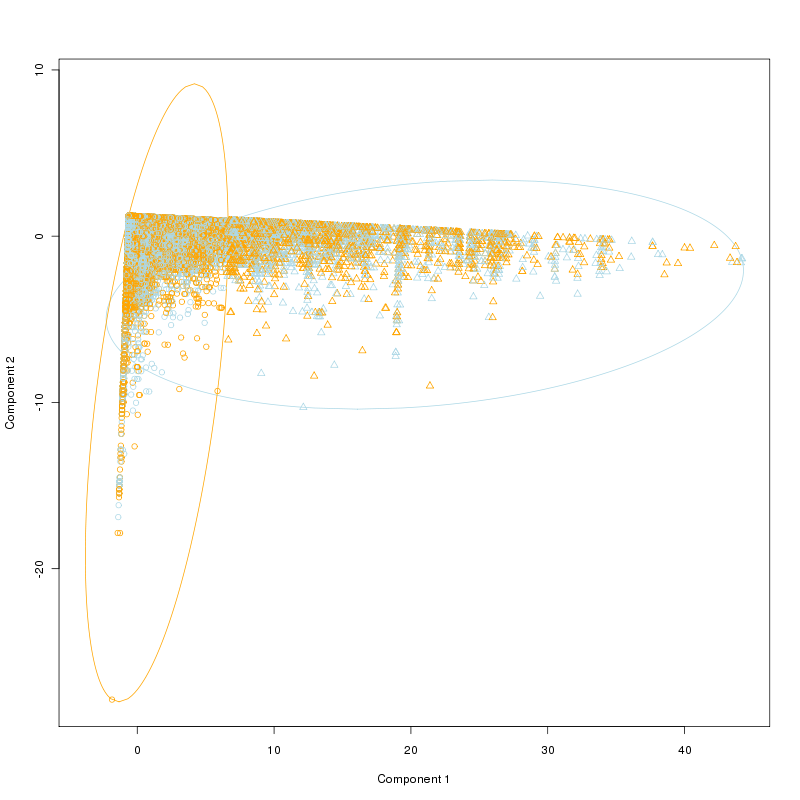}\label{fig:lcg-iat_rt-cluster-clara-part}}
\quad
\quad
\subfloat[The silhouette plot.]{\includegraphics[scale=0.25]{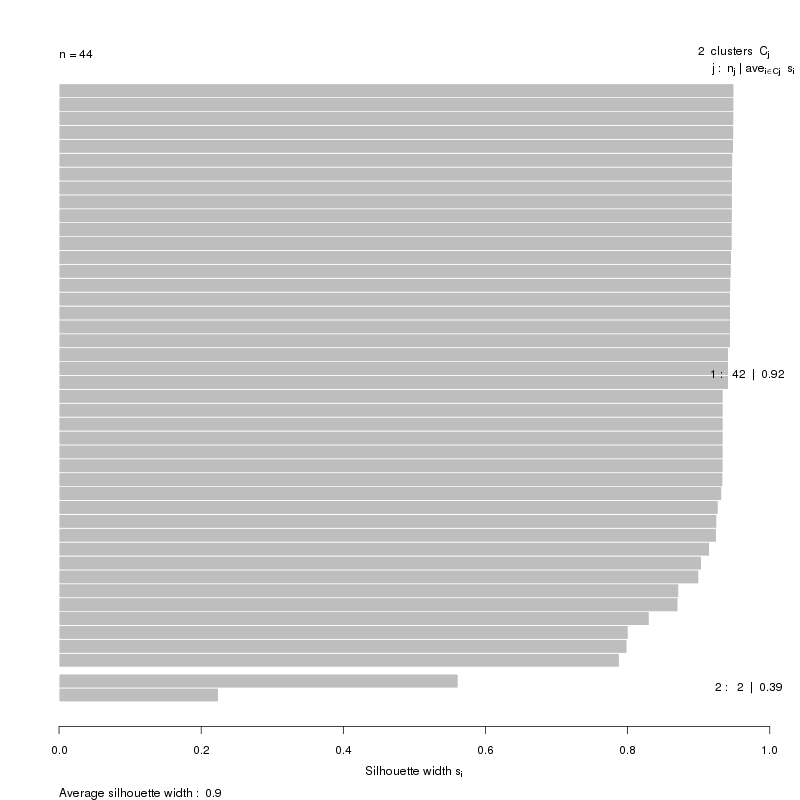}\label{fig:lcg-iat_rt-cluster-clara-sil}}
\caption{Clustering with CLARA: the CLUSPLOT and the silhouette plot of the $2$ clusters.}
\label{fig:lcg-iat_rt-cluster-clara}
\end{figure}

From the above results, we decide to use clusters obtained from CLARA since this algorithm is the only one that we are able to run on the whole data set and, at the same time, it was supported by a well-proven clustering validation measure.
In \mgFigRef{lcg-iat_rt-cluster-clara} are illustrated the CLUSPLOT \cite{Pison1999Clusplot} of the $2$ clusters obtained with CLARA (see \mgFigRef{lcg-iat_rt-cluster-clara-part}) as well as the related silhouette plot (see \mgFigRef{lcg-iat_rt-cluster-clara-sil}).
The first cluster $C_1$ contains $183459$ observations and has its centroid at point $(3, 2325)$, while the second cluster $C_2$ contains $4061$ observations and has its centroid at point $(3, 188938)$.

\begin{figure}
\centering
\subfloat[Comparison between the real and synthetic EDF.]{\includegraphics[scale=0.25]{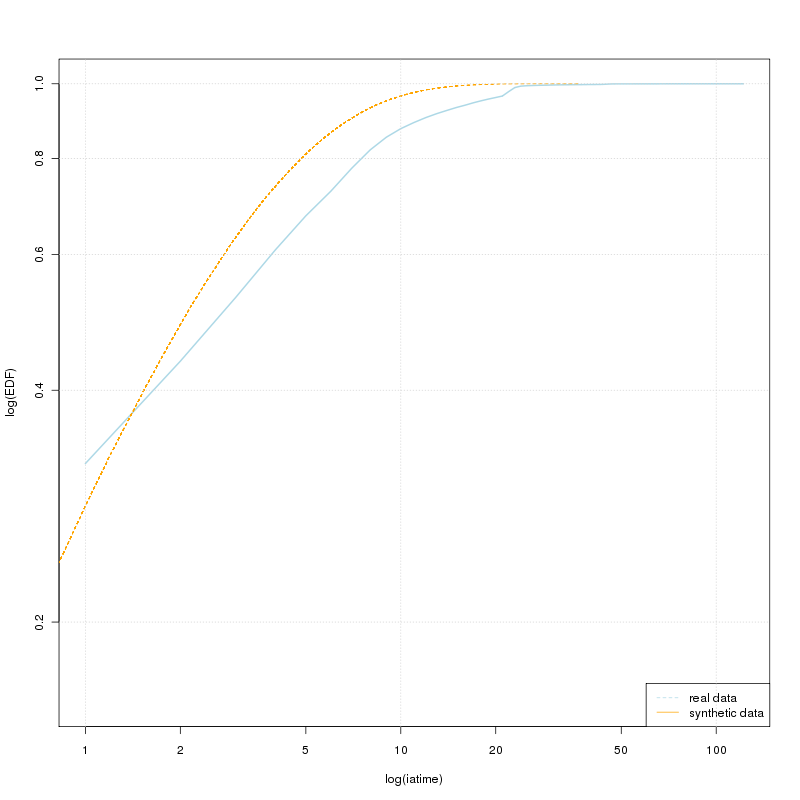}\label{fig:lcg-cmp-iat-model-cdf}}
\quad
\quad
\subfloat[Comparison between the real and synthetic CEDF.]{\includegraphics[scale=0.25]{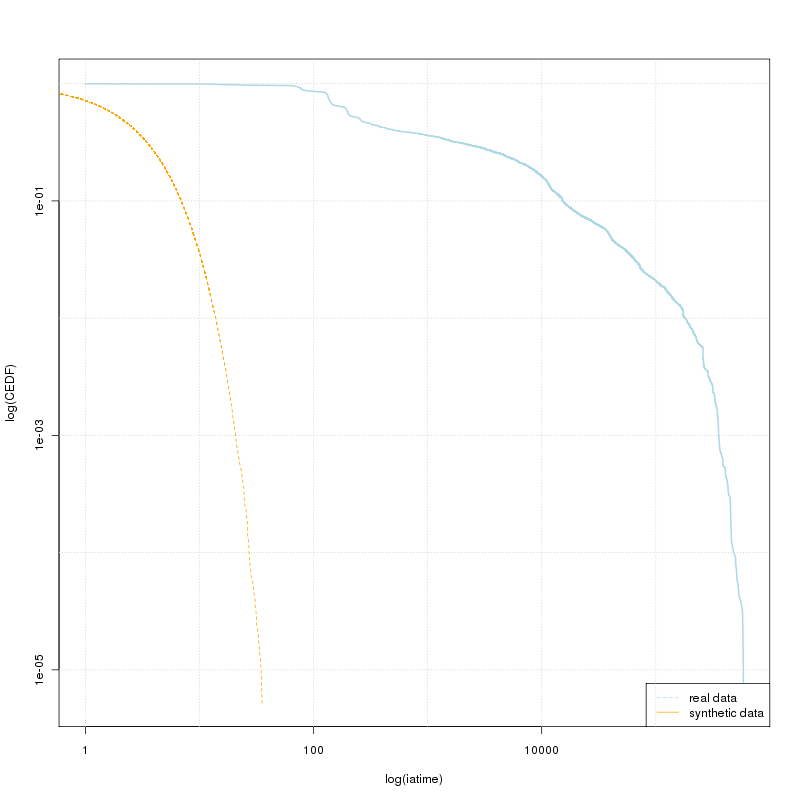}\label{fig:lcg-cmp-iat-model-ccdf}}
\caption{Comparison between real and synthetic workload for the \texttt{iatime} attribute.}
\label{fig:lcg-cmp-iat-model}
\end{figure}

\begin{figure}
\centering
\subfloat[Comparison between the real and synthetic EDF.]{\includegraphics[scale=0.25]{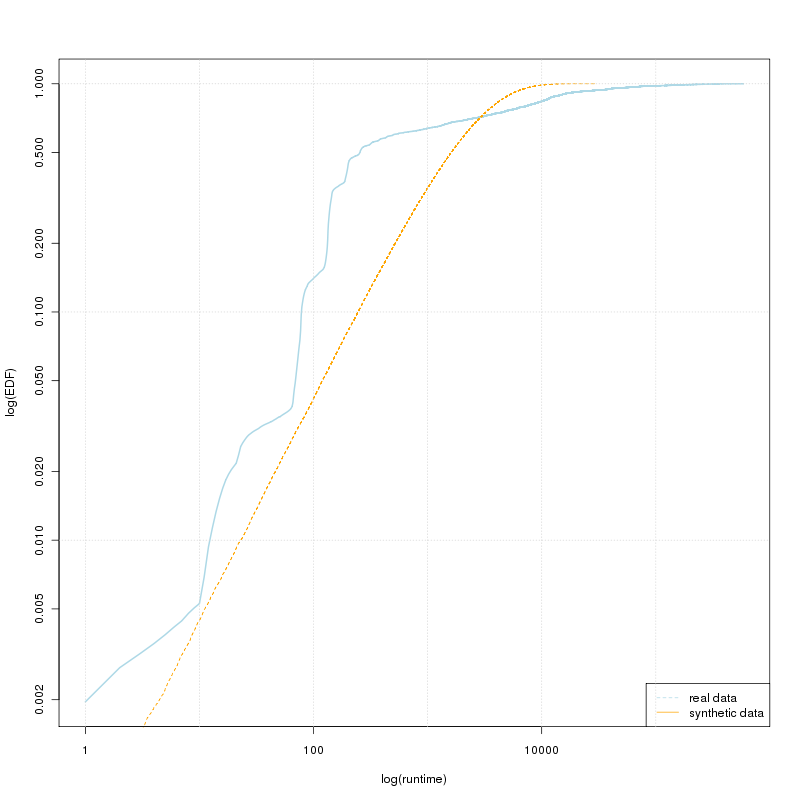}\label{fig:lcg-cmp-rt-model-cdf}}
\quad
\quad
\subfloat[Comparison between the real and synthetic CEDF.]{\includegraphics[scale=0.25]{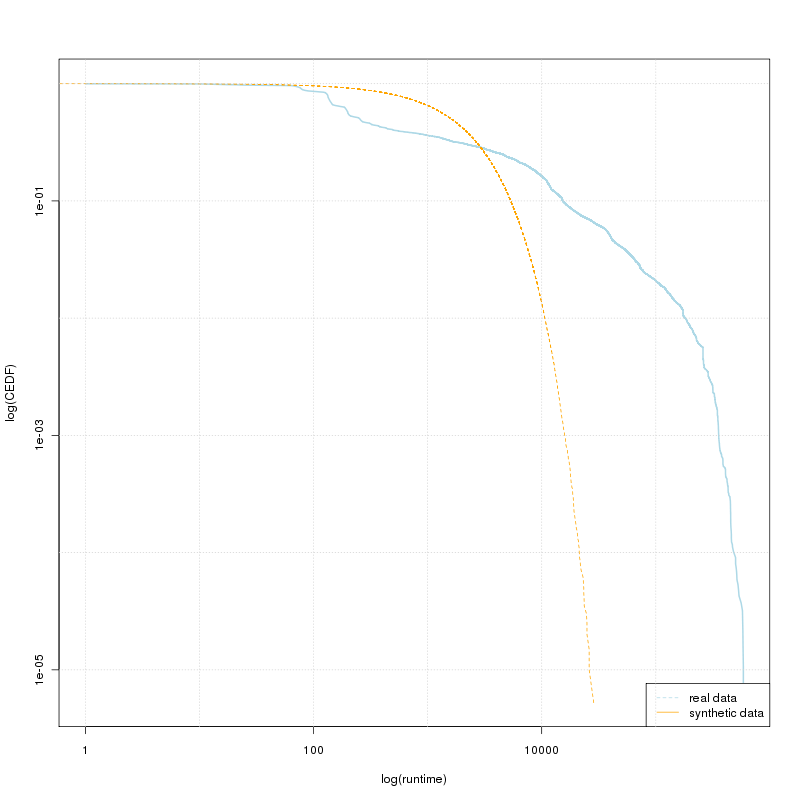}\label{fig:lcg-cmp-rt-model-ccdf}}
\caption{Comparison between real and synthetic workload for the \texttt{runtime} attribute.}
\label{fig:lcg-cmp-rt-model}
\end{figure}

Using the clustering obtained with CLARA, we create a model $\mathcal{M}_{LCG}$ for the LCG trace.
In order to validate our model, we run a simulation for creating a synthetic workload of $200000$ entries (by means of \mgAlgRef{synthetic-workload}), and compared, for each attribute, the empirical cumulative function of the real data with the one resulting from the generated data.
The result for the \texttt{iatime} attribute is shown in \mgFigRef{lcg-cmp-iat-model}, while that for the \texttt{runtime} attribute is reported in \mgFigRef{lcg-cmp-rt-model}.
For the \texttt{iatime} attribute we can observe that the model fails to reproduce the real tail behavior (see \mgFigRef{lcg-cmp-iat-model-ccdf}), while it is able to capture the general behavior of the body part of the distribution (see \mgFigRef{lcg-cmp-iat-model-cdf}).
For what concerns the \texttt{runtime} attribute we can note that the model fails to mimic the real tail behavior too (see \mgFigRef{lcg-cmp-rt-model-ccdf}) even if to a less extent than the \texttt{iatime} attribute, while it is able to capture the general behavior of the body part of the distribution (see \mgFigRef{lcg-cmp-rt-model-cdf}), even if it is unable to model the multi-modality of the real workload due to the use of univariate distributions.

\section{Conclusion} \label{sec:conclusion}

In this work, we model workload characteristics of real grid systems through data mining techniques and propose and algorithm to generate synthetic workload traces.

Specifically, we apply cluster analysis as a tool for summarizing workload characteristics.
We choose to use partitional clustering for representing workload characteristics by means of possibly few representatives.
Since every cluster algorithm impose a structure that is inherent to the nature of the algorithm itself, we evaluate, before running any cluster algorithm, the clustering tendency of the data set.
Then we apply three famous cluster algorithms: CLARA, a prototype-based cluster algorithm suitable for large data sets, DBSCAN, a density-based cluster algorithm, and MCLUST, a model-based clustering algorithm based on multivariate Gaussian fitting.
By means of clustering validation measures, we find that the algorithm that provided the best result was CLARA.
On the basis of the result obtained with CLARA, we create a workload model by mixing cluster information and Bayesian probability.
Finally, we validate our model through the generation of synthetic workload, where interarrival time and runtime variates were generated according to an Exponential distribution with mean equal to the centroid of the clusters.

The result of model validation indicates that the model fails, in general, to reproduce the behavior of the tail of the empirical distribution.
This is almost surely due to the use of the Exponential distribution for modeling both the body and the tail of the empirical distribution.
As a matter of fact, during the statistical analysis of workload characteristics we find an indication of power-law behavior for which the Exponential distribution is not a good model candidate.
As a final remark, we want to point out that our analysis was greatly influenced by the curse of the dimensionality problem.
Indeed, we are unable to determine clustering tendency on the whole data set, nor we are able to run MCLUST on the entire data set.
For these reasons, possible future works include the evaluation of different cluster algorithms, more suitable for large data set and, at the same time, more reliable than CLARA.
Another interesting aspect that would be worth investigating would be the comparison of the model presented in this work with the one obtained with traditional distribution fitting.
Finally, it could be interesting to evaluate different probability distributions, other than the Exponential distribution, possibly taking care of the power-law behavior.


\bibliographystyle{IEEEtran}
\bibliography{GridWorkloadDM}

\begin{thebibliography}{10}
\providecommand{\url}[1]{#1}
\csname url@samestyle\endcsname
\providecommand{\newblock}{\relax}
\providecommand{\bibinfo}[2]{#2}
\providecommand{\BIBentrySTDinterwordspacing}{\spaceskip=0pt\relax}
\providecommand{\BIBentryALTinterwordstretchfactor}{4}
\providecommand{\BIBentryALTinterwordspacing}{\spaceskip=\fontdimen2\font plus
\BIBentryALTinterwordstretchfactor\fontdimen3\font minus
  \fontdimen4\font\relax}
\providecommand{\BIBforeignlanguage}[2]{{%
\expandafter\ifx\csname l@#1\endcsname\relax
\typeout{** WARNING: IEEEtran.bst: No hyphenation pattern has been}%
\typeout{** loaded for the language `#1'. Using the pattern for}%
\typeout{** the default language instead.}%
\else
\language=\csname l@#1\endcsname
\fi
#2}}
\providecommand{\BIBdecl}{\relax}
\BIBdecl

\bibitem{Foster2001Anatomy}
I.~Foster, C.~Kesselman, and S.~Tuecke, ``The anatomy of the {G}rid: Enabling
  scalable {V}irtual {O}rganizations,'' \emph{International Journal of High
  Performance Computer Application}, vol.~15, no.~3, pp. 200--222, 2001.

\bibitem{Foster-2004-Grid2}
I.~Foster and C.~Kesselman, Eds., \emph{The Grid 2: Blueprint for a New
  Computing Infrastructure}, 2nd~ed., ser. The Morgan Kaufmann Series in
  Computer Architecture and Design.\hskip 1em plus 0.5em minus 0.4em\relax
  Burlington: Morgan Kaufmann, 2004.

\bibitem{Garey1979Computers}
M.~R. Garey and D.~S. Johnson, \emph{Computers and {I}ntractability; A Guide to
  the Theory of {NP}-{C}ompleteness}.\hskip 1em plus 0.5em minus 0.4em\relax W.
  H. Freeman \& Co., January 1979.

\bibitem{Shapiro1991Knowledge}
G.~Piatetsky-Shapiro, ``Knowledge discovery in real databases: A report on the
  ijcai-89 workshop,'' \emph{AI Magazine}, vol.~11, no.~5, pp. 37--54, January
  1991.

\bibitem{Fayyad1996DataOverview}
U.~M. Fayyad, G.~Piatetsky-Shapiro, and P.~Smyth, ``From {D}ata {M}ining to
  {K}nowledge {D}iscovery: An overview,'' in \emph{Advances in Knowledge
  Discovery and Data Mining}.\hskip 1em plus 0.5em minus 0.4em\relax Menlo
  Park, CA, USA: Association for the Advancement of Artificial Intelligence,
  1996, pp. 1--34.

\bibitem{Montgomery2002Applied}
D.~C. Montgomery and G.~C. Runger, \emph{Applied Statistics and Probability for
  Engineers}, 3rd~ed.\hskip 1em plus 0.5em minus 0.4em\relax John Wiley \&
  Sons, Inc., 2002.

\bibitem{Newman2005PowerLaws}
\BIBentryALTinterwordspacing
M.~E.~J. Newman, ``Power laws, {P}areto distributions and {Z}ipf's law,''
  \emph{Contemporary Physics}, vol.~46, no.~5, pp. 323--351, September 2005.
  [Online]. Available: \url{http://arxiv.org/abs/cond-mat/0412004}
\BIBentrySTDinterwordspacing

\bibitem{Brockwell2002Introduction}
P.~J. Brockwell and R.~A. Davis, \emph{Introduction to Time Series and
  Forecasting}, 2nd~ed.\hskip 1em plus 0.5em minus 0.4em\relax Springer-Verlag,
  2002.

\bibitem{Beran1994Statistics}
J.~Beran, \emph{Statistics for Long-Memory Processes}.\hskip 1em plus 0.5em
  minus 0.4em\relax Chapman \& Hall/CRC, October 1994.

\bibitem{Embrechts2000Introduction}
P.~Embrechts and M.~Maejima, ``An introduction to the theory of self-similar
  stochastic processes,'' \emph{International Journal of Modern Physics B},
  vol.~14, no. 12/13, pp. 1399--1420, May 2000.

\bibitem{Adler1998Practical}
R.~J. Adler, R.~E. Feldman, and M.~S. Taqqu, Eds., \emph{A Practical Guide to
  Heavy Tails. Statistical Techniques and Applications}.\hskip 1em plus 0.5em
  minus 0.4em\relax Birkh{\"a}user, 1998.

\bibitem{Tukey1977EDA}
J.~Tukey, \emph{Exploratory Data Anlysis}.\hskip 1em plus 0.5em minus
  0.4em\relax Addison-Wesley, 1977.

\bibitem{Wilk1968Probability}
M.~B. Wilk and R.~Gnanadesikan, ``Probability plotting methods for the analysis
  of data,'' \emph{Biometrika}, vol.~55, no.~1, pp. 1--17, 1968.

\bibitem{Crovella2001Performance}
M.~Crovella, ``Performance evaluation with heavy tailed distributions,'' in
  \emph{Proc. of the 7th International Workshop on Job Scheduling Strategies
  for Parallel Processing (JSSPP '01)}.\hskip 1em plus 0.5em minus 0.4em\relax
  Springer-Verlag, 2001, pp. 1--10.

\bibitem{Feitelson2006Metrics}
D.~G. Feitelson and D.~Tsafrir, ``Metrics for mass-count disparity,'' in
  \emph{Proc. of the 14th IEEE International Symposium on Modeling, Analysis,
  and Simulation of Computer and Telecommunication Systems (MASCOTS
  2006)}.\hskip 1em plus 0.5em minus 0.4em\relax IEEE Computer Society, Sep
  2006, pp. 61--68.

\bibitem{Lorenz1905Methods}
M.~O. Lorenz, ``Methods of measuring the concentration of wealth,'' vol.~9,
  no.~70, pp. 209--219, 1905.

\bibitem{Chambers1983Graphical}
J.~M. Chambers, W.~Cleveland, B.~Kleiner, and P.~Tukey, \emph{Graphical Methods
  for Data Analysis}, ser. The Wadsworth statistics/probability series.\hskip
  1em plus 0.5em minus 0.4em\relax Duxbury Press, 1983.

\bibitem{Box1970Time}
G.~E.~P. Box and G.~M. Jenkins, \emph{Time Series Analysis: Forecasting and
  Control}.\hskip 1em plus 0.5em minus 0.4em\relax Holden-Day, 1970.

\bibitem{Tan2006Introduction}
P.-N. Tan, M.~Steinbach, and V.~Kumar, \emph{Introduction to Data
  Mining}.\hskip 1em plus 0.5em minus 0.4em\relax Addison-Wesley, 2006.

\bibitem{Feitelson2007Locality}
D.~G. Feitelson, ``Locality of sampling and diversity in parallel system
  workloads,'' in \emph{Proc. of the 21st annual international conference on
  Supercomputing (ICS '07)}.\hskip 1em plus 0.5em minus 0.4em\relax Seattle,
  Washington, USA: ACM, 2007, pp. 53--63.

\bibitem{Kaufman1990Finding}
L.~Kaufman and P.~J. Rousseeuw, \emph{Finding Groups in Data: An Introduction
  to Cluster Analysis}.\hskip 1em plus 0.5em minus 0.4em\relax
  Wiley-Interscience, 1990.

\bibitem{Fraley1999MCLUST}
C.~Fraley and A.~E. Raftery, ``{MCLUST}: Software for model-based cluster
  analysis,'' \emph{Journal of Classification}, vol.~16, no.~2, pp. 297--306,
  1999.

\bibitem{Ester1996Density}
M.~Ester, H.-P. Kriegel, J.~Sander, and X.~Xu, ``A density-based algorithm for
  discovering clusters in large spatial databases with noise,'' in \emph{Proc.
  of the 2nd International Conference on Knowledge Discovery and Data Mining
  (KDD '96)}.\hskip 1em plus 0.5em minus 0.4em\relax Portland, OR, USA: AAAI
  Press, December 1996, pp. 226--231.

\bibitem{Dempster1977Maximum}
A.~P. Dempster, N.~M. Laird, and D.~B. Rubin, ``Maximum likelihood from
  incomplete data via the {EM} algorithm,'' \emph{Journal of the Royal
  Statistical Society. Series B (Methodological)}, vol.~39, no.~1, pp. 1--38,
  1977.

\bibitem{Schwarz1978Estimating}
G.~Schwarz, ``Estimating the dimension of a model,'' \emph{The Annals of
  Statistics}, vol.~6, no.~2, pp. 461--464, 1978.

\bibitem{Ling1973Computer}
R.~L. Ling, ``A computer generated aid for cluster analysis,''
  \emph{Communications of the ACM}, vol.~16, no.~6, pp. 355--361, 1973.

\bibitem{Bezdek2002VAT}
J.~C. Bezdek and R.~J. Hathaway, ``{VAT}: A tool for visual assessment of
  (cluster) tendency,'' in \emph{Proc. of the 2002 International Joint
  Conference on Neural Networks (IJCNN '02)}, vol.~3, Honolulu, HI, USA, 2002,
  pp. 2225--2230.

\bibitem{Park2009Visualisation}
L.~A.~F. Park, J.~C. Bezdek, and C.~A. Leckie, ``Visualisation of clusters in
  very large rectangular dissimilarity data,'' in \emph{Proc. of the 4th
  International Conference on Autonomous Robots and Agents}, G.~S. Gupta and
  S.~C. Mukhopadhyay, Eds., February 2009, pp. 251--256.

\bibitem{Jain1988Algorithms}
A.~K. Jain and R.~C. Dubes, \emph{Algorithms for Clustering Data}.\hskip 1em
  plus 0.5em minus 0.4em\relax Prentice-Hall, Inc., 1988.

\bibitem{Hopkins1954New}
B.~Hopkins and J.~G. Skellam, ``A new method for determining the type of
  distribution of plant individuals,'' \emph{Annals of Botany}, vol.~18, pp.
  213--227, 1954.

\bibitem{Zeng1985Comparison}
G.-Z. Zeng and R.~C. Dubes, ``A comparison of tests for randomness,''
  \emph{Pattern Recognition}, vol.~18, no.~2, pp. 191--198, 1985.

\bibitem{Fernandez2000Improved}
J.~A.~F. Pierna and D.~L. Massart, ``Improved algorithm for clustering
  tendency,'' \emph{Analytica Chimica Acta}, vol. 408, no. 1--2, pp. 13--20,
  2000.

\bibitem{Rousseeuw1987Silhouettes}
P.~J. Rousseeuw, ``Silhouettes: a graphical aid to the interpretation and
  validation of cluster analysis,'' \emph{Journal of Computational and Applied
  Mathematics}, vol.~20, no.~1, pp. 53--65, 1987.

\bibitem{Cern_LCG_Url}
C.~E. pour~la Recherche Nucl{\'e}aire~(CERN), ``{L}arge {H}adron {C}ollider
  {C}omputing {G}rid ({LCG}),'' http://lcg.web.cern.ch/LCG/.

\bibitem{Li2006Modeling}
H.~Li, M.~Muskulus, and L.~Wolters, ``Modeling job arrivals in a data-intensive
  grid,'' in \emph{Proc. of the 12th International Workshop on Job Scheduling
  Strategies for Parallel Processing (JSSPP)}.\hskip 1em plus 0.5em minus
  0.4em\relax Saint Malo, France: Springer-Verlag, Jun 2006, pp. 210--231.

\bibitem{Feitelson_PWA_Url}
D.~Feitelson, ``{P}arallel {W}orkload {A}rchive ({PWA}),''
  http://www.cs.huji.ac.il/labs/parallel/workload/, visited on March 3rd.

\bibitem{dcsr}
M.~Guazzone, ``dcsr-workload-characterization,'' Online:
  \url{https://github.com/sguazt/dcsr-workload-characterization}, 2007.

\bibitem{Pison1999Clusplot}
G.~Pison, A.~Struyf, and P.~J. Rousseeuw, ``Displaying a clustering with
  {CLUSPLOT},'' \emph{Computational Statistics \& Data Analysis}, vol.~30,
  no.~4, pp. 381--392, 1999.

\end{thebibliography}

\end{document}